\documentclass[lettersize,journal]{IEEEtran}
\usepackage[noadjust]{cite}
\usepackage{amsmath,amsfonts}
\usepackage{algorithmic}
\usepackage{array}
\usepackage[caption=false,font=normalsize,labelfont=sf,textfont=sf]{subfig}
\usepackage{textcomp}
\usepackage{stfloats}
\usepackage{url}
\usepackage{verbatim}
\usepackage{graphicx}
\usepackage{multirow}
\usepackage{multicol}
\usepackage{url}
\usepackage[dvipsnames]{xcolor}
\usepackage[linesnumbered,algoruled,boxed,lined]{algorithm2e}

\usepackage{layouts}

\usepackage{tikz}
\usepackage{hyperref}
\usepackage{lipsum}
\newcommand\copyrighttext{%
  \footnotesize \textcopyright \textcopyright 2024 IEEE. Personal use of this material is permitted. Permission from IEEE must be obtained for all other uses, in any current or future media, including reprinting/republishing this material for advertising or promotional purposes, creating new collective works, for resale or redistribution to servers or lists, or reuse of any copyrighted component of this work in other works.
  DOI: \href{https://ieeexplore-1ieee-1org-1000047su005e.wbg2.bg.agh.edu.pl/document/10528859}{10.1109/TASLP.2024.3399618}}
\newcommand\copyrightnotice{%
\begin{tikzpicture}[remember picture,overlay]
\node[anchor=south,yshift=5pt] at (current page.south) {\fbox{\parbox{\dimexpr\textwidth-\fboxsep-\fboxrule\relax}{\copyrighttext}}};
\end{tikzpicture}%
}

\def\BibTeX{{\rm B\kern-.05em{\sc i\kern-.025em b}\kern-.08em
    T\kern-.1667em\lower.7ex\hbox{E}\kern-.125emX}}
\usepackage{balance}
\begin{document}

\IEEEaftertitletext{\vspace{-0.2cm}}

\title{On Ambisonic Source Separation with Spatially Informed Non-negative Tensor Factorization}
\author{Mateusz Guzik, \emph{Student Member, IEEE} and Konrad Kowalczyk, \emph{Senior Member, IEEE}

\thanks{This work was supported in part by the National Science Centre, Poland, under Grant 2017/25/B/ST7/01792 and Grant 2021/42/E/ST7/00452, and in part by the Excellence Initiative — Research University Program for the AGH University of Krakow. We gratefully acknowledge Poland’s high performance Infrastructure PLGrid (ACK Cyfronet AGH) for providing computer facilities and support within computational Grant PLG/2023/016177 and Grant PLG/2023/016371. The authors are with the Signal Processing Group, Institute of Electronics, Faculty of Computer Science, Electronics and Telecommunications, AGH University of Krakow, 30-059 Krakow, Poland (e-mail: mguzik@agh.edu.pl; konrad.kowalczyk@agh.edu.pl).}}

\markboth{Published in IEEE/ACM TRANSACTIONS ON AUDIO, SPEECH AND LANGUAGE PROCESSING}{?}

\maketitle

\copyrightnotice

\newcommand{\prb}{\ensuremath{p}} 
\newcommand{\prm}{\ensuremath{\Theta}} 
\newcommand{\std}{\ensuremath{\sigma}} 
\newcommand{\N}{\ensuremath{\mathcal{N}}} 
\newcommand{\Nc}{\ensuremath{\mathcal{N}_c}} 
\newcommand{\W}{\ensuremath{\mathcal{W}}} 
\newcommand{\IW}{\ensuremath{\mathcal{IW}}} 
\newcommand{\brc}[1]{\ensuremath{\left(#1\right)}} 
\newcommand{\crBrc}[1]{\ensuremath{\left\{#1\right\}}} 
\newcommand{\sqBrc}[1]{\ensuremath{\left[#1\right]}} 
\newcommand{\vtBrc}[1]{\ensuremath{\left|#1\right|}} 
\newcommand{\conj}[1]{\ensuremath{#1^\ast}} 
\newcommand{\hrm}[1]{\ensuremath{#1^\mathrm{H}}} 
\newcommand{\tps}[1]{\ensuremath{#1^\mathrm{T}}} 
\newcommand{\diag}[1]{\ensuremath{\mathrm{diag}\crBrc{#1}}} 
\newcommand{\trc}[1]{\ensuremath{\mathrm{tr}\brc{#1}}} 
\newcommand{\sqFrNrm}[1]{\ensuremath{\left\lVert#1\right\rVert^2_\mathrm{F}}} 
\newcommand{\exptd}[1]{\ensuremath{\mathrm{E} \crlBrc{#1}}} 
\newcommand{\inv}[1]{\ensuremath{#1^{-1}}} 
\newcommand{\sq}[1]{\ensuremath{#1^{2}}} 
\newcommand{\pow}[2]{\ensuremath{#1^{#2}}} 
\newcommand{\freqIdx}{\ensuremath{f}} 
\newcommand{\freqNum}{\ensuremath{F}} 
\newcommand{\enFreqInd}{\ensuremath{\freqIdx = 1, \ldots, \freqNum}} 

\newcommand{\frameIdx}{\ensuremath{t}} 
\newcommand{\frameNum}{\ensuremath{T}} 
\newcommand{\enFrameInd}{\ensuremath{\frameIdx = 1, \ldots, \frameNum}} 

\newcommand{\micIdx}{\ensuremath{i}} 
\newcommand{\micNum}{\ensuremath{I}} 
\newcommand{\enMicInd}{\ensuremath{\micIdx = 1, \ldots, \micNum}} 

\newcommand{\srcIdx}{\ensuremath{j}} 
\newcommand{\srcNum}{\ensuremath{J}} 
\newcommand{\enSrcInd}{\ensuremath{\srcIdx = 1, \ldots, \srcNum}} 

\newcommand{\shdIdx}{\ensuremath{l}} 
\newcommand{\shdNum}{\ensuremath{L}} 
\newcommand{\enShdInd}{\ensuremath{\shdIdx = 1, \ldots, \shdNum}} 

\newcommand{\dirIdx}{\ensuremath{d}} 
\newcommand{\dirNum}{\ensuremath{D}} 
\newcommand{\enDirInd}{\ensuremath{\dirIdx = 1, \ldots, \dirNum}} 

\newcommand{\cmpIdx}{\ensuremath{k}} 
\newcommand{\cmpNum}{\ensuremath{K}} 
\newcommand{\enCmpInd}{\ensuremath{\cmpIdx = 1, \ldots, \cmpNum}} 
\newcommand{\col}{\ensuremath{\theta}} 
\newcommand{\az}{\ensuremath{\phi}} 
\newcommand{\rad}{\ensuremath{r}} 
\newcommand{\micCol}{\ensuremath{\col_\micIdx}} 
\newcommand{\micAz}{\ensuremath{\az_\micIdx}} 
\newcommand{\srcCol}{\ensuremath{\theta_\srcIdx}} 
\newcommand{\srcAz}{\ensuremath{\phi_\srcIdx}} 
\newcommand{\micAngs}{\ensuremath{\Phi}} 
\newcommand{\srcAngs}{\ensuremath{\Omega}} 
\newcommand{\sndSpeed}{\ensuremath{c}} 
\newcommand{\radFreq}{\ensuremath{\omega}} 
\newcommand{\wavNum}{\ensuremath{\kappa}} 
\newcommand{\sh}{\ensuremath{Y}} 
\newcommand{\radCoeff}{\ensuremath{b}} 
\newcommand{\smplWghts}{\ensuremath{\alpha_\micIdx}} 
\newcommand{\shOrd}{\ensuremath{n}} 
\newcommand{\maxShOrd}{\ensuremath{N}} 
\newcommand{\shDeg}{\ensuremath{m}} 
\newcommand{\legPoly}{\ensuremath{\mathcal{P}_\shOrd^\shDeg}} 
\newcommand{\BMat}{\ensuremath{\mathbf{B}_\freqIdx}} 
\newcommand{\BinvMat}{\ensuremath{\mathbf{B}_\freqIdx^{-1}}} 
\newcommand{\strVec}{\ensuremath{\mathbf{y}}} 
\newcommand{\strMat}{\ensuremath{\mathbf{Y}}} 
\newcommand{\pres}{\ensuremath{P}} 
\newcommand{\shPres}{\ensuremath{P}} 
\newcommand{\shdSigVec}{\ensuremath{\mathbf{p}_{\freqIdx \frameIdx}}} 
\newcommand{\shdCmpSigVec}{\ensuremath{\mathbf{a}_{\freqIdx \frameIdx}}} 
\newcommand{\cmpSig}{\ensuremath{A}} 
\newcommand{\mcSigVec}{\ensuremath{\Tilde{\mathbf{a}}_{\freqIdx \frameIdx}}} 
\newcommand{\empMcCov}{\ensuremath{\covMat_{\Tilde{\mathbf{a}}, \freqIdx \frameIdx}}} 
\newcommand{\mdlMcCov}{\ensuremath{\hat{\covMat}_{\Tilde{\mathbf{a}}, \freqIdx \frameIdx}}} 
\newcommand{\strv}{\ensuremath{\mathbf{y}}} 
\newcommand{\srcSig}{\ensuremath{S_{\srcIdx \freqIdx \frameIdx}}} 
\newcommand{\srcSigVec}{\ensuremath{\mathbf{s}_{\freqIdx \frameIdx}}} 
\newcommand{\mcSrcSig}{\ensuremath{\Tilde{S}_{\srcIdx \freqIdx \frameIdx}}} 
\newcommand{\estSrcVec}{\ensuremath{\widehat{\mathbf{s}}_{\srcIdx \freqIdx \frameIdx}}} 
\newcommand{\covMat}{\ensuremath{\mathbf{R}}} 
\newcommand{\micCov}{\ensuremath{\covMat_{\freqIdx \frameIdx}}}
\newcommand{\SCM}{\ensuremath{\mathbf{\Xi}_\srcIdx}}
\newcommand{\mdlCov}{\ensuremath{\widehat{\covMat}_{\freqIdx \frameIdx}}}
\newcommand{\mdlSrcCov}{\ensuremath{\widehat{\covMat}_{\srcIdx \freqIdx \frameIdx}}}
\newcommand{\estSCM}{\ensuremath{\widehat{\mathbf{\Xi}}_\srcIdx}}
\newcommand{\dirCov}{\ensuremath{\mathbf{\Sigma}_\dirIdx}}
\newcommand{\dirSCM}{\ensuremath{\mathbf{\Phi}_\srcIdx}}
\newcommand{\micMcCov}{\ensuremath{\widetilde{\covMat}_{\freqIdx \frameIdx}}}
\newcommand{\idtyMat}{\ensuremath{\mathbf{I}}} 
\newcommand{\micCovEU}{\ensuremath{\widetilde{\covMat}_{\freqIdx \frameIdx}}}
\newcommand{\micCovIS}{\ensuremath{\micCov}}
\newcommand{\Qjk}{\ensuremath{Q_{\srcIdx \cmpIdx}}}
\newcommand{\Wfk}{\ensuremath{W_{\freqIdx \cmpIdx}}}
\newcommand{\Htk}{\ensuremath{H_{\frameIdx \cmpIdx}}}
\newcommand{\Zjd}{\ensuremath{Z_{\srcIdx \dirIdx}}}
\newcommand{\Vjft}{\ensuremath{V_{\srcIdx \freqIdx \frameIdx}}}
\newcommand{\estVjft}{\ensuremath{\widehat{V}_{\srcIdx \freqIdx \frameIdx}}}
\newcommand{\frDeg}{\ensuremath{\nu}}
\newcommand{\WScale}{\ensuremath{\mathbf{\Psi}_\srcIdx^\W}}
\newcommand{\IWScale}{\ensuremath{\mathbf{\Psi}_\srcIdx^\IW}}
\newcommand{\latCmpI}{\ensuremath{\mathbf{C}_{\srcIdx \freqIdx \frameIdx \cmpIdx \dirIdx}}}
\newcommand{\latCmpII}{\ensuremath{\mathbf{U}_{\freqIdx \frameIdx}}}
\newcommand{\EUAuxFnc}{\ensuremath{\mathcal{L}^\text{EU}}}
\newcommand{\ISAuxFnc}{\ensuremath{\mathcal{L}^\text{IS}}}
\newcommand{\AuxFnc}{\ensuremath{\mathcal{L}}}

\begin{abstract}
This article presents a Non-negative Tensor Factorization based method for sound source separation from Ambisonic microphone signals.
The proposed method enables the use of prior knowledge about the Directions-of-Arrival (DOAs) of the sources, incorporated through a constraint on the Spatial Covariance Matrix (SCM) within a Maximum a Posteriori (MAP) framework.
Specifically, this article presents a detailed derivation of four algorithms that are based on two types of cost functions, namely the squared Euclidean distance and the Itakura-Saito divergence, which are then combined with two prior probability distributions on the SCM, that is the Wishart and the Inverse Wishart.
The experimental evaluation of the baseline Maximum Likelihood (ML) and the proposed MAP methods is primarily based on first-order Ambisonic recordings, using four different source signal datasets,  three with musical pieces and one containing speech utterances.
We consider under-determined, determined, as well as over-determined scenarios by separating two, four and six sound sources, respectively. Furthermore, we evaluate the proposed algorithms for different spherical harmonic orders and at different reverberation time levels, as well as in non-ideal prior knowledge conditions, for increasingly more corrupted DOAs.
Overall, in comparison with beamforming and a state-of-the-art separation technique, as well as the baseline ML methods, the proposed MAP approach offers superior separation performance in a variety of scenarios, as shown by the analysis of the experimental evaluation results, in terms of the standard objective separation measures, such as the SDR, ISR, SIR and SAR.
\end{abstract}


\section{Introduction}
\noindent
The field of spatial sound processing has recently been greatly influenced by the Ambisonic technology, owing to its useful qualities, which often make it the preferred choice for audio recording and reconstruction \cite{zotter2019ambisonics, frank2015producing}.
In particular, the recording procedure based on the Spherical Harmonic (SH) decomposition enables uniform coverage of the entire sound scene, which makes Ambisonics very well suited for spatial audio in general.
Moreover, since the knowledge about the composition of the sound field is acquired, rather than the specific values of sound pressure in a fixed microphone setup, the Ambisonic technology also enables to conveniently decouple the recording and reconstruction stages.
This quality is useful with respect to easily-scalable and versatile immersive audio playback systems, because the decoder can be straightforwardly determined for various loudspeaker setups.
As a result, the interest in the Ambisonic audio format for 360 audio-video, augmented and virtual reality \cite{olivieri2019scene}, has attracted the attention of many researchers.

The classical Ambisonic linear decoding suffers from low-order limitations, arising as a consequence of an insufficient spatial resolution, which may cause a notable degradation in perceived sound quality \cite{bernschutz2014binaural, avni2013spatial, bertet2013investigation, solvang2008spectral}.
This has led to the formulation of alternative, more advanced, signal-dependent methods that involve parametric representation of the sound field \cite{kowalczyk2015parametric}, such as Directional Audio Coding (DirAC) \cite{pulkki2007spatial} and Higher‐Order Directional Audio Coding (HO-DirAC) \cite{politis2015sector}, High Angular Resolution Planewave Expansion (HARPEX) \cite{barrett2010new} or COding and Multidirectional Parameterization of Ambisonic Sound Scenes (COMPASS) \cite{politis2018compass}.
Depending on the adopted model, these signal-dependent methods describe the sound field with various parameters, which need to be estimated from the Spherical Harmonic Domain (SHD) signals and are later utilized during the reconstruction stage.
Among different parameters, the source power spectral densities and the individual source signals are of key importance, since the accuracy of their estimation determines the level of audio quality and control over the reproduction of the spatial sound scene.
One other important parameter, which often appears in the context of spatial sound and its reproduction, are the Directions of Arrival (DOAs) associated with the positions of sound sources relative to the array position.

Apart from the classical methods, such as fixed and adaptive spatial filtering \cite{rafaely2015fundamentals, teutsch2007modal}, an effort in terms of the methods for sound source separation in the SHD has already been observed.
This includes the deep learning based approaches \cite{lluis2023direction, herzog2022ambisep, perotin2018multichannel, nishiguchi2018dnn}, the Non-negative Tensor Factorization (NTF) based methods \cite{guzik2022ntf, mitsufuji2020multichannel, hafsati2019sound, nikunen2018multichannel}, and many other \cite{munoz2021ambisonics, varanasi2018stochastic, kalkur2015joint, epain2012independent}.
While it is well established in the literature that the prior DOA knowledge can be utilized to substantially raise the quality of source separation \cite{duong2011acoustically, duong2013spatial, duong2013raport, fras2021maximum}, only a limited literature concerns the usage of prior DOA information in the SHD for sound source separation\cite{guzik2021incorporation, guzik2022wishart, lluis2023direction}.

Concerning methods that operate directly on the microphone signals, rather than in the SHD, there exist a multitude of approaches to sound source separation \cite{sekiguchi2019fast, sawada2019review, duong2010under} and some of them incorporate prior localization knowledge \cite{fras2021maximum, duong2013spatial}.
While these methods could potentially be applied directly to the raw microphone signals, as opposed to the encoded Ambisonic recordings, the existing research indicates that an enhanced performance can be achieved by performing separation in the SHD \cite{hafsati2019sound}.
Moreover, direct application of the sound source separation methods, formulated for the classical microphone arrays, is only possible if these methods do not assume prior knowledge about the microphone array, they are not direction-aware, i.e. the form of the steering vector is unknown.
Unfortunately, in the case of the prior DoA incorporating methods, the steering vectors must be defined explicitly.
While in the classical array setup, the steering vector is defined as a phase shift between the microphones, the Ambisonic signals are not characterized by a phase shift.
This prevents the direct usage of the already existing prior DoA incorporating methods with the Ambisonic signals, unless they are explicitly adapted to the SHD.
On the other hand, the methods formulated for the classical microphone arrays have to adopt frequency-dependent steering vectors, as opposed to the case of Ambisonic signals.
By utilization this frequency-independence, broadband estimation is seamlessly enabled, owing to the fact that the Ambisonic SCMs can be averaged over frequency bands.
This can often result in an enhanced performance, while at the same time enabling significant computational speedup, because the update equations for the spatial components can be calculated once, as opposed to calculating them independently per each frequency band.
This, connected with the so far limited interest in incorporation of prior DoA information for Ambisonic source separation, constitutes the motivation underlying this research.

In this work, we develop on our conference publication \cite{guzik2022wishart}, in which we presented an extension of an NTF-based algorithm from \cite{nikunen2018multichannel}.
In its basic form, this algorithm employs a Parallel Factor Analysis (PARAFAC) NTF for the source spectrogram estimation \cite{cichocki2009nonnegative}, while the spatial information is encoded through a SHD Spatial Covariance Matrix (SCM) model.
Within this SCM model, the individual source SCMs are approximated as a weighted sum of SH DOA kernels, which results in an additional NTF parameter, referred to as a spatial selector.

In this article, we rewrite the baseline, originally formulated as a Maximum Likelihood (ML) problem, using the Maximum a Posteriori (MAP) framework, such that the prior DOA knowledge is incorporated through a prior probability distribution on the SCM, which guides the estimation of the spatial properties.
We present a detailed derivation including two cost functions, namely the squared Euclidean distance and the Itakura-Saito divergence, and we combine them with two prior probability distributions on SCM, namely the Wishart and the Inverse Wishart.
Altogether, we introduce and evaluate four new MAP variants of the baseline algorithm and a new MAP\textrightarrow{}ML algorithm that is a sequential combination of the proposed MAP and the baseline ML solutions.
The experimental evaluation is primarily based on first-order Ambisonic recordings.
To account for the variety of source signals encountered in real-life applications, we generate four different datasets, three of which include musical pieces and one containing speech utterances.
We consider the cases of two, four and six sound sources, since this enables the investigation of the performance in over-determined, determined, and under-determined scenarios, when using first-order Ambisonics, derived from 4 SH signals.
Furthermore, we evaluate the proposed algorithms for different spherical harmonic orders and at different reverberation time levels.
Finally, we evaluate the proposed algorithms for an increasingly more corrupted DOAs and we propose a combined MAP\textrightarrow{}ML solution to overcome the deterioration observed in separation performance.
The objective evaluation is based on measures widely used for source separation, namely the Signal-to-Distortion-Ratio, Image-to-Spatial-Distortion-Ratio, Signal-to-Interference-Ratio and Signal-to-Artifacts-Ratio \cite{vincent2007first}.
The results of experimental evaluation clearly indicate the advantages and the overall superior separation performance of the proposed MAP solutions, with respect to the baseline ML approaches.

Overall, the novelties reported in this article include two spatial localization priors formulated for the Ambisonic SCM model, namely the Wishart and the Inverse Wishart, and their application to the likelihood cost functions based on the squared Euclidean distance and the Itakura-Saito divergence.
As a result, five novel MAP algorithms for source separation in the spherical harmonic domain are introduced, four of which are obtained as a combination of the proposed localization priors with the considered cost functions, while the additional fifth algorithm is a MAP\textrightarrow{}ML combination of one of the proposed MAP solutions and the baseline ML.
For all of the introduced algorithms we include a detailed derivation of the update equations.
Note that one of the presented algorithms, namely the spatial localization prior with Wishart distribution applied to the Euclidean squared distance, was preliminary presented in our conference paper \cite{guzik2022wishart}, in which the normalization factor introduced herein was not included in the updates and only a limited experimental evaluation was performed.
 In the present journal article, however, we perform an in-depth experimental evaluation of all five new algorithms, followed by an extensive discussion of the results and their impact on practical applications.
\section{Spherical Harmonic Domain}\label{sec:spherical_harmonic_domain}
\noindent
Ambisonic signals are tyically obtained via the spherical Fourier transform of the recordings made using multi-microphone spherical setups, which in general enable to obtain spherical harmonics of high orders. For the first-order Ambisonics, four directional microphones can alternatively be used \cite{zotter2019ambisonics}. Since in this work we address the general problem of source separation in the SHD of any order, we include the description of the more general higher-order Ambisonic procedure which involves the spherical Fourier transform.
Note however, that regardless of the way the Ambisonic signals are obtained, our method and derivations are valid.

Let us consider an array consisting of \enMicInd{} omnidirectional microphones arranged on a sphere with radius \rad{}.
The origin of the coordinate system is located in the geometrical center of the array, while the colatitude and the azimuth angles associated with the placement of the sensors are, respectively, denoted by
\(
    \micCol{}
    \in
    [0, \pi)
\)
and
\(
    \micAz{}
    \in
    [0, 2 \pi)
    ,
\)
with the colatitude equal to 0 and $2 \pi$, respectively, at the north and south poles.
For a given frequency f, corresponding to the wave number
\(
    \wavNum = 2 \pi \mathrm{f} / \sndSpeed
    ,
\)
where \sndSpeed{} is the sound wave velocity, the acoustic pressure captured by the microphones \pres \brc{\wavNum, \frameIdx, \rad, \micAz, \micCol} at time frame $\frameIdx{}=1,\ldots,T$ and its SHD representation
\(
    \shPres_\shOrd^\shDeg \brc{\wavNum, \frameIdx, \rad}
\)
can be determined given the approximation of the spherical Fourier transform and its inverse as \cite{teutsch2007modal}

\begin{equation}
    \shPres_\shOrd^\shDeg \brc{\wavNum, \frameIdx, \rad}
    =
    \sum_\micIdx^\micNum
    \smplWghts
    \pres \brc{\wavNum, \frameIdx, \rad, \col_\micIdx, \az_\micIdx}
    \sh_\shOrd^\shDeg \brc{\col_\micIdx, \az_\micIdx}
    ,
\end{equation}
\begin{equation}
    \pres \brc{\wavNum, \frameIdx, \rad, \col_\micIdx, \az_\micIdx}
    =
    \sum_{\shOrd=0}^\maxShOrd
    \sum_{\shDeg=-\shOrd}^\shOrd
    \shPres_\shOrd^\shDeg \brc{\wavNum, \frameIdx, \rad}
    \sh_\shOrd^\shDeg \brc{\col_\micIdx, \az_\micIdx}
    ,
\end{equation}
where the weights \smplWghts{} are introduced to support the approximated equality, while their values depend on the adopted spatial sampling scheme \cite{rafaely2004analysis}.
The truncation order, or the upper boundary of approximation, is given by
\(\shOrd = 0, \ldots, \maxShOrd,\) which results in
\(
    \enShdInd = \brc{\maxShOrd + 1}^2
\)
SH signals, while the number of microphones required to obtain the desired order depends on the adopted spatial sampling scheme \cite{rafaely2015fundamentals, rafaely2004analysis}.
Concerning Ambisonics, the real-valued spherical harmonic
\(
    \sh_\shOrd^\shDeg
\)
of order \shOrd{} and degree \shDeg{} is defined as
\begin{multline}
    \sh_\shOrd^\shDeg \left( \col, \az \right)
    = \sqrt{
            [\brc{\shOrd - \vtBrc{\shDeg}}!]^{-1}
            \brc{2\shOrd+1}
            \brc{\shOrd + \vtBrc{\shDeg}}!
    }\;\\
    \mathcal{P}_\shOrd^{\vtBrc{\shDeg}}
    \Big(\sin \brc{\col}\Big)
    y_\shDeg \brc{\az},
\end{multline}
\begin{equation}
    y_\shDeg \brc{\az}
    =
    \begin{cases}
        \sqrt{2} \sin \brc{\vtBrc{\shDeg} \az}, & \shDeg<0\\
        1, & \shDeg=0\\
        \sqrt{2} \cos \brc{\shDeg \az}, & \shDeg>0 
    \end{cases}
    ,
\end{equation}
\cite{pulkki2018parametric} (Sec. 6.2, page 144), where
\(
    \shOrd \ge \vtBrc{\shDeg}
\)
and
\(
    \mathcal{P}_\shOrd^{\vtBrc{\shDeg}} \brc{ \cdot }
\) is the associated Legendre polynomial \cite{teutsch2007modal} (Sec. B3, page 216).

Assuming static point sources, the sound pressure on the surface of a spherical microphone array can be expressed in the SHD in the following matrix form
\begin{equation}\label{eq:shd_signal}
    \shdSigVec
    =
    \BMat
    \strMat
    \srcSigVec
    ,
\end{equation}
where \enFreqInd{} denotes the frequency indices.
Omitting \rad{} for brevity, the SH signals are arranged in a vector according to
\(
    \shdSigVec
    =
    \tps{
        \sqBrc{
                \shPres_0^0 \brc{\wavNum, \frameIdx},
                \shPres_1^{-1} \brc{\wavNum, \frameIdx},
                \shPres_1^0 \brc{\wavNum, \frameIdx},
                \shPres_1^1 \brc{\wavNum, \frameIdx},
                \ldots,
                \shPres_\maxShOrd^\maxShOrd \brc{\wavNum, \frameIdx}
            }
        }
    \in
    \mathbb{C}^{\shdNum \times 1}
    ,
\)
while
\(
    \srcSigVec
    =
    \tps{
        \sqBrc{
            S_{1ft},
            \ldots,
            S_{\srcNum ft}
        }
    }
    \in 
    \mathbb{C}^{\srcNum \times 1}
\)
is the vector with complex-valued elements of source signal spectra, where \enSrcInd{} denotes the number of sound sources.
In \eqref{eq:shd_signal}, the modal matrix
\(
    \BMat{} \in \mathbb{C}^{\shdNum \times \shdNum}
\) is given by a diagonal matrix
\(
    \BMat = \mathrm{diag} \brc{\sqBrc{\radCoeff_0 \brc{\wavNum \rad}, \radCoeff_1 \brc{\wavNum \rad}, \radCoeff_1 \brc{\wavNum \rad}, \ldots \radCoeff_\maxShOrd \brc{\wavNum \rad}}}
    ,
\)
where the radial coefficients
\(
    \radCoeff_\shOrd \brc{\wavNum \rad},
\)
which lie on the main diagonal, depend on array configuration, since they model the microphone array frequency response.
In cases of commonly used geometries, there exist analytical expressions for the radial coefficients, e.g. for open and rigid sphere configurations \cite{teutsch2007modal}.
The steering matrix
\(
    \strMat{} \in \mathbb{R}^{\shdNum \times \srcNum}
\) is composed of \srcNum{} steering vectors
\(
    \strVec_\srcIdx \in \mathbb{R}^{\shdNum \times 1}
\)
that point towards the
directions associated with the positions of sound sources, as given by
\begin{equation}
    \strMat \brc{\srcAngs}
    =
    \begin{bmatrix}
        \sh_0^0 \brc{\vartheta_1, \varphi_1} &
        \dots &
        \sh_0^0 \brc{\vartheta_\srcNum, \varphi_\srcNum} \\
        \sh_1^{-1} \brc{\vartheta_1, \varphi_1} &
        \dots &
        \sh_1^{-1} \brc{\vartheta_\srcNum, \varphi_\srcNum} \\
        \sh_1^0 \brc{\vartheta_1, \varphi_1} &
        \dots &
        \sh_1^0 \brc{\vartheta_\srcNum, \varphi_\srcNum} \\
        \sh_1^1 \brc{\vartheta_1, \varphi_1} &
        \{\dots &
        \sh_1^1 \brc{\vartheta_\srcNum, \varphi_\srcNum} \\
        \vdots &
        \ddots &
        \vdots \\
        \sh_\maxShOrd^\maxShOrd \brc{\vartheta_1, \varphi_1} &
        \dots &
        \sh_\maxShOrd^\maxShOrd \brc{\vartheta_\srcNum, \varphi_\srcNum} \\
    \end{bmatrix}
    ,
\end{equation}
where 
\(
    \srcAngs
    =
    \crBrc{\brc{\vartheta_1, \varphi_1}, \ldots, \brc{\vartheta_\srcNum, \varphi_\srcNum}}
\)
is a set of colatitude
\(
    \vartheta_\srcIdx
\)
and azimuth
\(
    \varphi_\srcIdx
\)
angles, expressing jointly the DOA of the \srcIdx{}-th sound source.

For further processing, the frequency- and angle-dependent components are often decoupled by multiplying \eqref{eq:shd_signal} by \BinvMat{} from the left side, arriving at the SH plane wave signal model
\begin{equation}\label{eq:SHD_plane_wave}
    \shdCmpSigVec = \strMat \, \srcSigVec
    ,
\end{equation}
where
\(
    \shdCmpSigVec
    =
    \BinvMat
    \shdSigVec
\)
and
\(
    \shdCmpSigVec \in \mathbb{C}^{\shdNum \times 1}
    .
\)

Assuming statistical independence between source signals, which imposes zero mutual cross-correlation, the microphone covariance matrix for the SH plane wave signal model \eqref{eq:SHD_plane_wave} is expressed as a sum of contributions from individual sound sources, as
\begin{equation}\label{eq:SH_plane_wave_micCov}
    \micCov
    =
    \mathrm{E}
    \crBrc{
    \shdCmpSigVec
    \hrm{\shdCmpSigVec}}
    =
    \mathrm{E}
    \crBrc{
    \strMat
    \srcSigVec
    \hrm{\srcSigVec}
    \tps{\strMat}
    }
    =
    \sum_\srcIdx^\srcNum
    \Vjft
    \strVec_\srcIdx
    \tps{\strVec_\srcIdx}
    ,
\end{equation}
where $\mathrm{E}$ $\crBrc{\cdot}$ is the expectation operator,
\(
    \Vjft
    =
    \srcSig
    \conj{\srcSig}
    \in \mathbb{R}
\)
denotes the signal variance for the \srcIdx{}-th source,
\(
    \hrm{\brc{\cdot}}
\)
denotes the Hermitian conjugate transpose and
\(
    \conj{\brc{\cdot}}
\)
is the complex conjugate operation.
\section{Ambisonic NTF with spatial covariance matrix model}\label{sec:III}
\noindent
Since the covariance matrix of the microphone signals is a source of both spectral and spatial information, it commonly appears in context of multichannel audio analysis.
For instance, in \cite{nikunen2018multichannel} the microphone covariance matrix is expressed using the Non-negative Tensor Factorization with a Spatial Covariance Matrix model, for source separation with Ambisonic signals.

The SH microphone covariance matrix \micCov{} is approximated by \srcNum{} source variances
\(
    \estVjft \in \mathbb{R}
\)
and their corresponding SCMs
\(
\estSCM \in \mathbb{R}^{\shdNum \times \shdNum}
\)
as given by
\begin{equation}\label{eq:micCovModel}
    \micCov
    \approx
    \mdlCov
    =
    \sum_\srcIdx^\srcNum
    \mdlSrcCov
    =
    \sum_\srcIdx^\srcNum
    \estVjft
    \estSCM
    .
\end{equation}

Further, the source variances are subject to the NTF, which decomposes them into a set of representative spectral patterns and the corresponding activation weights. 
In the context of audio, the spectral patterns can be thought of as single notes or phoneme-like speech fragments, although without any explicit conditioning there is no guarantee as to their actual content.
On the other hand, the time activation weights contain information concerning the spectral patterns occurrences, meaning time and intensity of their observation.
For more details on the general application of NTF to audio source separation please refer to \cite{ozerov2018introduction}.
In this work, we adopt the following Parallel Factor Analysis \cite{cichocki2009nonnegative} model, proven effective for sound source separation \cite{ozerov2018introduction, nikunen2018multichannel}, given as
\begin{equation}\label{eq:NTF}
    \estVjft
    =
    \sum_\cmpIdx^\cmpNum
    \Qjk
    \Wfk
    \Htk
    ,
\end{equation}
where, \Wfk{} contains the spectral patterns, \Qjk{} maps them to sound sources and \Htk{} consists of the time activation weights, while \cmpNum{} denotes an arbitrary number of components.

Considering sound source separation under reverberant conditions, the rank-1 covariance model \mdlSrcCov{} of \eqref{eq:SH_plane_wave_micCov} based on a single anechoic steering vector would not be suitable to accurately represent such a complex sound field \cite{duong2010under}.
Therefore, to handle reverberant conditions, the SCMs are expressed using multiple SH DOA kernels \dirCov{}, weighted with the so-called spatial selector \Zjd{}, as
\begin{equation}\label{eq:SCM}
    \estSCM
    =
    \sum_\dirIdx^\dirNum
    \Zjd
    \dirCov
    .
\end{equation}
The direction-dependent kernels
\(
    \dirCov \in \mathbb{R}^{\shdNum \times \shdNum}
\)
are fixed and need to be defined for a sufficiently high number of discrete directions \enDirInd{} distributed uniformly on a spherical manifold, where the minimum number of directions depends on the adopted spatial sampling scheme \cite{rafaely2004analysis}.
Additionally, in order to remove the ambiguity between the source- and the direction-dependent magnitudes, the spatial selector needs to be constrained, such that
\(
    \sum_\dirIdx^\dirNum \Zjd = 1
    .
\)
For further information concerning the NTF indeterminacies, kindly refer to \cite{ozerov2009multichannel}.

By substituting \eqref{eq:NTF} and \eqref{eq:SCM} into \eqref{eq:micCovModel}, the final model of the SHD microphone covariance matrix is obtained as
\begin{equation}\label{eq:final_model}
    \mdlCov
    =
    \sum_\srcIdx^\srcNum
    \sum_\cmpIdx^\cmpNum
    \Qjk
    \Wfk
    \Htk
    \sum_\dirIdx^\dirNum
    \Zjd
    \dirCov
    ,
\end{equation}
where 
\(
    \prm
    =
    \crBrc{\Qjk, \Wfk, \Htk, \Zjd}
\)
forms a set of parameters to be estimated.
\section{Maximum likelihood probabilistic models}\label{sec:4}
\noindent
Parameters
\(
    \prm
    =
    \crBrc{\Qjk, \Wfk, \Htk, \Zjd}
\)
can be estimated by optimization of a suitable likelihood function.
In this work, we investigate two probabilistic models, which are associated with the most commonly utilized cost functions for the NTF based source separation \cite{sawada2013multichannel, fevotte2009nonnegative}, namely the Squared Euclidean distance and the Itakura-Saito divergence.
The State-of-the-art (SOTA) Maximum Likelihood based algorithms formulated in the SHD, such as  \cite{nikunen2018multichannel, munoz2021ambisonics, mitsufuji2020multichannel}, are commonly based on these two types of the cost functions.

\subsection{Squared Euclidean distance}\label{subsec:EU}
\noindent
Since the squared Euclidean distance (EU) is a scale-sensitive cost function \cite{fevotte2009nonnegative}, it is often preferred to model the magnitude spectrogram rather than the power spectrogram, as the influence of certain observations might be unnecessarily enhanced \cite{sawada2013multichannel}.
Therefore, similarly to \cite{guzik2022ntf, guzik2022wishart, nikunen2018multichannel, sawada2013multichannel}, we compress the magnitude of the observed microphone signals according to
\(
    \mcSigVec
    =
    \vtBrc{\shdCmpSigVec}^{-1/2}
    \odot\
    \shdCmpSigVec
    ,
\)
where
\(
    \vtBrc{\cdot}^{-1/2}
\)
denotes the entry-wise inverse of the square root of the magnitude, while \(\odot\) is the Hadamard product.
With the empirical Ambisonic signal covariance matrix determined by 
\begin{equation}
    \micCovEU
    =
    \mathrm{E}
    \crBrc{
    \mcSigVec
    \hrm{\mcSigVec}
    }
    ,
\end{equation}
the values on its main diagonal contain the magnitude spectrum of the observed mixtures.

Assuming independence of individual observations, the following conditional probabilistic model can be adopted \cite{sawada2013multichannel}
\begin{equation}\label{eq:EU_model}
    \prb_{\text{EU}} \brc{\prm \Big| \micCovEU}
    =
    \prod
    \Nc \Bigl(\Bigl[\mdlCov\Bigr]_{l_1 l_2} \Big| \Bigl[\micCovEU\Bigr]_{l_1 l_2}, \sq{\std_{\text{EU}}}\Bigr)
    ,
\end{equation}
where \Nc{} denotes the complex Gaussian distribution,
\(
    \sqBrc{\cdot}_{l_1 l_2}
\)
denotes the entry-wise matrix indexing and
\(
    \sq{\std_{\text{EU}}}
\)
is the variance of the complex Gaussian distribution.
Please note, that whenever we use a product \(\prod\) or summation \(\sum\) symbols without explicitly specifying their ranges, it should be acknowledged that the respective operation is performed over all possible indices, such that a scalar value is produced.
Additionally, explicit dependencies on the set of parameters \(\Theta\) were omitted throughout the text to increase readability.
The negative log-likelihood for the model given by \eqref{eq:EU_model} has the form of the squared Euclidean distance 
\begin{multline}\label{eq:EU_nll}
    - \log \brc{\prb_{\text{EU}} \brc{\prm \Big| \micCovEU}}
    \stackrel{c}{=}\\
    \inv{\brc{\pi \sq{\std_{\text{EU}}}}}
    \sum
    \trc{\mdlCov \hrm{\mdlCov}}
    -2 \trc{\mdlCov \hrm{\micCovEU}}
    ,
\end{multline}
where
\(
    \stackrel{c}{=}
\)
denotes an equality up to a constant, allowing us to omit the additive elements that remain fixed throughout the estimation and therefore do not impact the final estimates.
This includes terms dependent solely on empirical covariance matrices or distribution parameters, that would otherwise disappear during differentiation.
Optimization of the negative log-likelihood \eqref{eq:EU_nll} with respect to model parameters \prm{} enables to estimate them in a ML sense.
Note that in \eqref{eq:EU_nll} the conjugate transpose operator \hrm{\brc{\cdot}} is included for the sake of completeness.
However, in further derivations, whenever possible, we omit it for brevity, as it is not necessary when dealing with Hermitian matrices, such as is the case in the considered equations.

One of the major advantages of the squared Euclidean distance is that the parameter update equations derived using the EU do not require a
\(\shdNum{} \times \shdNum{}\)
matrix inversion at each time-frequency bin, as will be presented in further parts of this article.
This is in contrast to the Itakura-Saito divergence in which a
\(\shdNum{} \times \shdNum{}\)
matrix inversion at each time-frequency bin is necessary.
The computational advantage of the squared Euclidean distance makes the EU well suited for Ambisonic applications, as the number of channels quickly grows with an increasing SH order, such that the dimensions of the SHD microphone covariance matrices are given by
\(4 \times 4\)
for
\(
    \maxShOrd = 1
    ,
\)
\(9 \times 9\)
for
\(
    \maxShOrd = 2
    ,
\)
\(16 \times 16\)
for
\(
    \maxShOrd = 3
    ,
\)
\(25 \times 25\)
for
\(
    \maxShOrd = 4
    ,
\)
respectively.
\subsection{Itakura-Saito divergence}\label{subsec:IS}
\noindent
Unlike the Squared Euclidean distance, the Itakura-Saito divergence (IS) is not scale-sensitive \cite{fevotte2009nonnegative}, therefore the empirical microphone covariance matrix is determined as
\(
    \micCov
    =
    \mathrm{E}
    \{
        \shdCmpSigVec
        \hrm{\shdCmpSigVec}
    \}
    ,
\)
without the magnitude compression.

Again, assuming independence of individual observations, we can adopt the conditional probabilistic model given by \cite{sawada2013multichannel}
\begin{multline}\label{eq:IS_prb} 
    \prb_{\text{IS}} \brc{\prm \Big| \micCovIS}
    =\\
    \prod
    \Nc \Bigl(\shdCmpSigVec \Big| \mathbf{0}, \mdlCov\Bigr)
    \inv{\sqBrc{\Nc \Bigl(\shdCmpSigVec \Big| \mathbf{0}, \micCovIS\Bigr)}},
\end{multline} 
which leads to the corresponding negative log-likelihood
\begin{equation}\label{eq:IS_nll}
    - \log \brc{\prb_{\text{IS}} \brc{\prm \Big| \micCovIS}}
    \stackrel{c}{=}
    \sum
    \trc{\micCovIS\inv{\mdlCov}}
    +
    \log \vtBrc{\mdlCov}.
\end{equation}

Although the IS is regarded as superior over the EU in classical NTF literature \cite{sawada2013multichannel, fevotte2009nonnegative}, the applicability of the IS to Ambisonic signals has so far been rather limited due to its aforementioned computational properties related to the occurrence of a matrix inverse in \eqref{eq:IS_nll}.
As will be presented in further parts of this article, within the update equations derived using the IS, an inversion of a \shdNum{}x\shdNum{} matrix is necessary in each time-frequency bin, making it less suitable for practical applications involving high-order spherical harmonics.
While literature notes the attempts to overcome this issue, e.g. by 
exploiting the Cholesky decomposition and proposing a parallel multi-architecture driver \cite{munoz2023efficient} or by joint diagonalization of the SCMs \cite{sekiguchi2019fast}, in this work we consider the IS in its basic form for simplicity, whereas application of the acceleration techniques into our methods is left for future work.
\section{Localization priors}\label{sec:5}
\noindent
In this article, we propose to incorporate the DOA information, associated with the position of a sound source, for the benefit of the aforementioned NTF based algorithm, through the SCM.
In order to constrain the SCM estimation, in later sections we reintroduce the problem within the MAP framework, by defining a suitable probabilistic prior distribution for the SCM.
In order to account for room reverberation, we employ two distributions presented in \cite{duong2013spatial, duong2013raport}, which are consistent with the theory of statistical room acoustics \cite{duong2011acoustically}, namely the Wishart and the Inverse Wishart.
While the formulation of the spatial localization prior with Wishart distribution in Sec. \ref{subsec:WLP} has already been preliminary shown in \cite{guzik2022wishart}, the formulation of the spatial localization prior with the Inverse Wishart distribution in Sec. \ref{subsec:IWLP} is presented for the first time.

\subsection{Wishart localization prior}\label{subsec:WLP}
\noindent
First, we define the prior probability
\(
    \prb_{\W} \brc{\prm}
    ,
\)
using the Wishart distribution \cite{maiwald2000calculation}
\begin{multline}
    \prb_{\W} \brc{\prm}
    =
    \prod
    \W \brc{\estSCM \Big| \WScale, \frDeg}
    =\\
    \prod
    \frac{
        \Bigl|\WScale\Bigr|^{-\frDeg}
        \Bigl|\estSCM\Bigr|^{\frDeg-\shdNum}
        e^{-\trc{\inv{\sqBrc{\WScale}} \estSCM}}
    }
    {
        \pi^{\shdNum \brc{\shdNum-1}/2}
        \prod_\shdIdx^\shdNum \Gamma \brc{\frDeg-\shdIdx+1}
    }
    ,
\end{multline}
where
\(
\WScale{} \in \mathbb{R}^{\shdNum \times \shdNum}
\)
is a Wishart scale matrix,
\(\frDeg{} \in \mathbb{R}\)
stands for the degrees of freedom and \(\Gamma(\cdot)\) is the gamma function.
The hyper-parameter \frDeg{} allows to control the acceptable deviation from the mean, while the density, its mean and variance are finite for
\(
    \frDeg > \shdNum - 1
    ,
\)
\(
    \frDeg > \shdNum
    ,
\)
and
\(
    \frDeg > \shdNum + 1,
\)
respectively.
Note that \frDeg{} does not necessarily have to be an integer value.
With the above, the negative log-prior is given by
\begin{multline}\label{eq:W_nlpr}
    - \log \brc{\prb_\W \brc{\prm}}
    \stackrel{c}{=}\\
    \sum
    \trc{\inv{\sqBrc{\WScale}} \estSCM}
    +
    \brc{\shdNum - \frDeg} \log \Bigl|\estSCM\Bigr|
    .
\end{multline}

Considering that the mean of the Wishart distribution is equal to
\(
    \frDeg \WScale
    ,
\)
we set the scale matrix to be
\(
    \WScale = \frac{1}{\frDeg} \brc{\strv_\srcIdx \hrm{\strv_\srcIdx} + \epsilon \idtyMat}
    ,
\)
based on the steering vectors for the known DOAs
\(
    \strv_\srcIdx
    ,
\)
the relative strength of the diffuse component
\(
    \epsilon
\)
and the diffuse component SH covariance matrix, which is simply an identity matrix \idtyMat{}.
However, we model different quantities within the two considered cost functions, i.e. magnitude vs. power spectrograms, in cases of the Squared Euclidean distance and the Itakura-Saito divergence, respectively.
In consequence, the diffuse component strength, and therefore the Wishart scale matrix, are defined differently for the Squared Euclidean distance and the Itakura-Saito divergence, as respectively given by
\begin{equation}\label{eq:WScale_EU}
    {\WScale}^\text{-EU}
    =
    \frac{1}{\frDeg}
    \dirSCM^\text{EU}
    =
    \frac{1}{\frDeg}
    \brc{
        \strv_\srcIdx
        \hrm{\strv_\srcIdx}
        +
        \epsilon^\text{EU}
        \idtyMat
    },
\end{equation} 
and
\begin{equation}\label{eq:WScale_IS}
    {\WScale}^\text{-IS}
    =
    \frac{1}{\frDeg}
    \dirSCM^\text{IS}
    =
    \frac{1}{\frDeg}
    \brc{
        \strv_\srcIdx
        \hrm{\strv_\srcIdx}
        +
        \epsilon^\text{IS}
        \idtyMat
    }.
\end{equation}
In the remaining parts of this work, we refer to the incorporation of prior localization knowledge through the Wishart distribution as Wishart Localization Prior (WLP).
\subsection{Inverse Wishart localization prior}\label{subsec:IWLP}
\noindent
In case of the Inverse Wishart distribution \cite{maiwald2000calculation}, the prior probability
\(
    \prb_{\IW} \brc{\prm}
\)
takes the following form 
\begin{multline}
    \prb_{\IW} \brc{\prm}
    =
    \prod
    \IW \brc{\estSCM \Big| \IWScale, \frDeg}
    =\\
    \prod
    \frac{
        \Bigl|\IWScale\Bigr|^\frDeg
        \Bigl|\estSCM\Bigr|^{-\brc{\shdNum+\frDeg}}
        e^{-\trc{\IWScale \inv{\estSCM}}}
    }
    {
        \pi^{\shdNum \brc{\shdNum-1}/2}
        \prod_\shdIdx^\shdNum \Gamma \brc{\frDeg-\shdIdx+1}
    }
    ,
\end{multline}
 which results in the corresponding negative log-prior
\begin{multline}\label{eq:IW_nlpr}
    - \log \brc{\prb_\IW \brc{\prm}}
    \stackrel{c}{=}\\
    \sum
    \trc{\IWScale \inv{\estSCM}}
    +
    \brc{\shdNum+\frDeg} \log \Bigl|\estSCM\Bigr|
    .
\end{multline}
Since the mean of the Inverse Wishart distribution is equal to
\(
    \inv{\brc{\frDeg - \shdNum}} \IWScale
    ,
\)
the scale matrices \IWScale{} for the Squared Euclidean distance and the Itakura-Saito divergence are respectively given by 
\begin{equation}\label{eq:IWScale_EU}
    {\IWScale}^\text{-EU}
    =
    \brc{\frDeg-\shdNum}
    \dirSCM^\text{-EU},
\end{equation}
and
\begin{equation}\label{eq:IWScale_IS}
    {\IWScale}^\text{-IS}
    =
    \brc{\frDeg-\shdNum}
    \dirSCM^\text{-IS}.
\end{equation}
Hereafter, we refer to the incorporation of prior localization knowledge through the Inverse Wishart distribution as Inverse Wishart Localization Prior (IWLP).
\section{Maximum a posteriori probabilistic models}\label{sec:MAP_models}
\noindent
According to the Bayes rule, the ML estimation can be reformulated as a MAP problem, by multiplying the likelihood with the prior probability and disregarding the marginal probability by treating it as a constant \cite{ozerov2018introduction}.
In this article, we propose the following four Maximum a Posteriori probabilistic models which are obtained by combining the two considered cost functions, namely the Squared Euclidean distance and the Itakura-Saito divergence, and the two considered localization priors, namely the Wishart and the Inverse Wishart distributions.
The presented models include the posterior probabilities for the squared Euclidean distance with Wishart Localization Prior and Inverse Wishart Localization Prior, respectively,
\begin{equation}\label{eq:EU_WLP_prb}
    \prb_{\text{EU-W}} \brc{\micCovEU \Big| \prm}
    \stackrel{c}{=}
    \prod
    \pow{\prb_{\text{EU}} \brc{\prm \Big| \micCovEU}}{1/\brc{\freqNum \frameNum}}
    \prb_{\text{W}} \brc{\prm},
\end{equation}
\begin{equation}\label{eq:EU_IWLP_prb}
    \prb_{\text{EU-IW}} \brc{\micCovEU \Big| \prm}
    \stackrel{c}{=}
    \prod
    \pow{\prb_{\text{EU}} \brc{\prm \Big| \micCovEU}}{1/\brc{\freqNum \frameNum}}
    \prb_{\text{IW}} \brc{\prm},
\end{equation}
and the Itakura-Saito divergence with Wishart Localization Prior and Inverse Wishart Localization Prior, respectively,
\begin{equation}\label{eq:IS_WLP_prb}
    \prb_{\text{IS-W}} \brc{\micCovIS \Big| \prm}
    \stackrel{c}{=}
    \prod
    \pow{\prb_{\text{IS}} \brc{\prm \Big| \micCovIS}}{1/\brc{\freqNum \frameNum}}
    \prb_{\text{W}} \brc{\prm},
\end{equation}
\begin{equation}\label{eq:IS_IWLP_prb}
    \prb_{\text{IS-IW}} \brc{\micCovIS \Big| \prm}
    \stackrel{c}{=}
    \prod
    \pow{\prb_{\text{IS}} \brc{\prm \Big| \micCovIS}}{1/\brc{\freqNum \frameNum}}
    \prb_{\text{IW}} \brc{\prm}.
\end{equation}
In order to assure a constant magnitude of prior influence with respect to the number of time-frequency bins, which translates to independence from the sampling rate and the recording length, posteriors \eqref{eq:EU_WLP_prb} - \eqref{eq:IS_IWLP_prb} are normalized by factor 
\(
    1/\brc{\freqNum \frameNum}
    ,
\)
which is equivalent to taking their geometric mean.
The estimation based on model \eqref{eq:EU_WLP_prb} was first introduced in \cite{guzik2022wishart}, however it did not involve the presented normalization, while models \eqref{eq:EU_IWLP_prb} - \eqref{eq:IS_IWLP_prb} constitute novel contributions.

\section{Derivation of update equations with squared Euclidean distance}\label{sec:EU_derivations}
\noindent
The update equations are derived by minimization of the appropriate negative log-likelihood or negative log-posterior, respectively, in the case of ML or MAP estimation.
Similarly to \cite{sawada2013multichannel}, in this work, we follow the majorization scheme \cite{de1994block, marshall1979inequalities} where a complex problem is simplified by defining suitable latent components and a corresponding auxiliary function.
Provided that certain conditions are met, this approach enables an indirect optimization of the original function.
In this section, we derive the final update equations for the presented algorithms based on the squared Euclidean distance cost function, in case of ML and MAP estimation with the Wishart and Inverse Wishart Localization Priors.
The update equations for the existing ML based algorithm \cite{nikunen2018multichannel} are presented in Sec. \ref{sec:VII_A}, while both MAP algorithms are derived in Secs. \ref{subsec:EU_WLP_derivation} and \ref{subsec:EU_IWLP_derivation}.
Note that as previously mentioned, parts of \ref{subsec:EU_WLP_derivation} were presented in \cite{guzik2022wishart}.

\subsection{Maximum likelihood}\label{sec:VII_A}
\noindent
To apply the majorization scheme to the negative log-likelihood of \eqref{eq:EU_nll}, we define the following latent components, with the corresponding auxiliary function
\begin{equation}\label{eq:lat_cmpI}
    \latCmpI
    =
    \inv{\mdlCov} \Qjk \Wfk \Htk \Zjd \dirCov
    ,
\end{equation}
\begin{multline}\label{eq:EU_aux_func}
    \EUAuxFnc
    =
    \inv{\brc{\pi \sq{\std_{\text{EU}}}}}
    \sum
    \sq{\Qjk} \sq{\Wfk} \sq{\Htk} \sq{\Zjd}
    \trc{\dirCov \inv{\latCmpI} \dirCov}\\
    -
    2 \Qjk \Wfk \Htk \Zjd \trc{\micCovEU \dirCov}
    ,
\end{multline}
respectively, such that
\(
    \sum_{\srcIdx, \cmpIdx, \dirIdx}^{\srcNum, \cmpNum, \dirNum}
    \latCmpI
    =
    \mathbf{I}
    ,
\)
while \latCmpI{} is a Hermitian positive definite matrix \cite{sawada2013multichannel}.
By following the proof presented in \cite{sawada2013multichannel}, it can be shown that minimization of the auxiliary function \eqref{eq:EU_aux_func} leads to an indirect minimization of the negative log-likelihood \eqref{eq:EU_nll}.

The iterative update equations are derived by differentiation of the auxiliary function \eqref{eq:EU_aux_func} with respect to the model parameters \prm{}, such that the latent components \eqref{eq:lat_cmpI} are considered to be fixed.
The resulting derivative is subject to the Multiplicative Update (MU) rule \cite{NIPS2000_f9d11525}, which is generalized by the following equation
\begin{equation}\label{eq:MU_rule}
    \lambda
    \xleftarrow[]{}
    \lambda
    \frac
    {\left[ \nabla_\lambda f \brc{\lambda} \right]_-}
    {\left[ \nabla_\lambda f \brc{\lambda} \right]_+}
    ,
\end{equation}
where the updated value of a parameter
\(\lambda\)
is calculated by multiplying its current value with the ratio of the absolute value of the negative to positive part of the gradient of a function
\(
    f \brc{\lambda}
\).
This form of update has the property, that once initialized with positive values, the estimated parameters retain their non-negativity \cite{NIPS2000_f9d11525}.

The partial derivative of the auxiliary function \eqref{eq:EU_aux_func} with respect to \Zjd{} takes the form of
\begin{multline}\label{eq:EU_Z_partial}
    \frac{\partial \EUAuxFnc}{\partial \Zjd}
    =
    \frac{2}{\pi \sq{\std_{\text{EU}}}}
    \sum_{\freqIdx, \frameIdx, \cmpIdx}^{\freqNum, \frameNum, \cmpNum}
    \sq{\Qjk}
    \sq{\Wfk}
    \sq{\Htk}
    \Zjd
    \trc{\dirCov \inv{\latCmpI} \dirCov}\\
    -
    \Qjk
    \Wfk
    \Htk
    \trc{\micCovEU \dirCov},
\end{multline}
while \eqref{eq:EU_Z_partial} subject to the MU rule \eqref{eq:MU_rule} results in the spatial selector update equation as given by
\begin{equation}\label{eq:EU_Z_MU}
    \Zjd
    \xleftarrow[]{}
    \Zjd
    \frac
    {
        \sum_{\freqIdx, \frameIdx}^{\freqNum, \frameNum}
        \estVjft \trc{\micCovEU \dirCov}
    }
    {
        \sum_{\freqIdx, \frameIdx}^{\freqNum, \frameNum}
        \estVjft \trc{\mdlCov \dirCov}
    }.
\end{equation}
Note that, as previously mentioned in Sec. \ref{sec:III}, the spatial selector needs to be re-scaled after each update \eqref{eq:EU_Z_MU} to uphold the unitary sum property.
Such re-scaling can be expressed by
\begin{equation}\label{eq:Z_MU_rescaling}
    \Zjd
    \xleftarrow[]{}
    \Zjd
    \inv{
        \sqBrc{
            \sum_\dirIdx^\dirNum \Zjd        
        }
    }
    .
\end{equation}
In case of the remaining model parameters, we provide the following parameter update equations
\begin{equation}\label{eq:EU_Q_MU}
    \Qjk
    \xleftarrow[]{}
    \Qjk
    \frac
    {
        \sum_{\freqIdx, \frameIdx}^{\freqNum, \frameNum}
        \Wfk \Htk \trc{\micCovEU \estSCM}
    }
    {
        \sum_{\freqIdx, \frameIdx}^{\freqNum, \frameNum}
        \Wfk \Htk \trc{\mdlCov \estSCM}
    },
\end{equation}
\begin{equation}\label{eq:EU_W_MU}
    \Wfk
    \xleftarrow[]{}
    \Wfk
    \frac
    {
        \sum_{\srcIdx, \frameIdx}^{\srcNum, \frameNum}
        \Qjk \Htk \trc{\micCovEU \estSCM}
    }
    {
        \sum_{\srcIdx, \frameIdx}^{\srcNum, \frameNum}
        \Qjk \Htk \trc{\mdlCov \estSCM}
    },
\end{equation}
\begin{equation}\label{eq:EU_H_MU}
    \Htk
    \xleftarrow[]{}
    \Htk
    \frac
    {
        \sum_{\srcIdx, \freqIdx}^{\srcNum, \freqNum}
        \Qjk \Wfk \trc{\micCovEU \estSCM}
    }
    {
        \sum_{\srcIdx, \freqIdx}^{\srcNum, \freqNum}
        \Qjk \Wfk \trc{\mdlCov \estSCM}
    },
\end{equation}
readily derived by application of an analogous procedure to that of the spatial selector \cite{nikunen2018multichannel}.
During runtime, update equations \eqref{eq:EU_Z_MU} - \eqref{eq:EU_H_MU} are sequentially repeated until convergence or some other stopping criterion.
\subsection{Maximum a posteriori with Wishart localization prior}\label{subsec:EU_WLP_derivation}
\noindent
In case of the MAP estimation with the probabilistic model of \eqref{eq:EU_WLP_prb}, we define an auxiliary function for the corresponding negative-log posterior using \eqref{eq:EU_aux_func} and \eqref{eq:W_nlpr},
as in \cite{guzik2022wishart}, by
\begin{equation}\label{eq:EU_W_aux}
    \AuxFnc^{\text{EU-W}}
    =\\
    \AuxFnc^\text{EU}
    - \log \brc{\prb_\text{W} \brc{\prm}}
    ,
\end{equation}
which is now the subject of differentiation, while the resulting derivatives are treated accordingly with the procedure described in Sec. \ref{sec:VII_A}.
Since the log-prior \eqref{eq:W_nlpr} does not depend on the spectral parameters, the localization prior does not directly affect the estimation of spectral parameters.
Therefore, the updates \eqref{eq:EU_Q_MU} - \eqref{eq:EU_H_MU} remain unchanged, while in this section, we consider only the spatial selector update equation.
The partial derivative of the auxiliary function \eqref{eq:EU_W_aux} with respect to \Zjd{} takes the following form
\begin{multline}\label{eq:EU_WLP_Z_partial}
    \frac{\partial \AuxFnc^{\text{EU-W}}}{\partial \Zjd}
    =
    \frac{\partial \EUAuxFnc}{\partial \Zjd}
    +
    \frac{\partial - \log \brc{\prb_\text{W} \brc{\prm}}}{\partial \Zjd}
    =
    \frac{\partial \EUAuxFnc}{\partial \Zjd}\\
    +
    \frDeg
    \trc{\inv{\sqBrc{\dirSCM^\text{EU}}} \dirCov}
    +
    \brc{\shdNum - \frDeg}
    \trc{\inv{\estSCM} \dirCov}
    ,
\end{multline}
where 
\(
\frac{\partial \EUAuxFnc}{\partial \Zjd}
\)
is defined in \eqref{eq:EU_Z_partial}. Subject to the MU rule of \eqref{eq:MU_rule}, the derivative of the auxiliary function \eqref{eq:EU_WLP_Z_partial} results in the spatial selector update equation as given by
\begin{multline}\label{eq:EU_WLP_Z_MU}
    \Zjd
    \xleftarrow[]{}
    \Zjd
    \Bigg[
        \sum_{\freqIdx, \frameIdx}^{\freqNum, \frameNum}
        \frac{\estVjft}{\freqNum \frameNum}
        \trc{\micCovEU \dirCov}
        +
        \frac{\pi \sq{\std_{\text{EU}}} \frDeg}{2}
        \trc{\inv{\estSCM} \dirCov}
    \Bigg]\\
    \inv{\Bigg[
        \sum_{\freqIdx, \frameIdx}^{\freqNum, \frameNum}
        \frac{\estVjft}{\freqNum \frameNum}
        \trc{\mdlCov \dirCov}
        +
        \frac{\pi \sq{\std_{\text{EU}}}}{2}
        \Big[
            \shdNum \trc{\inv{\estSCM} \dirCov}
            \\
            +
            \frDeg \trc{\inv{\sqBrc{\dirSCM^\text{EU}}} \dirCov}
        \Big]        
    \Bigg]}
    .
\end{multline}
\subsection{Maximum a posteriori with Inverse Wishart localization prior}\label{subsec:EU_IWLP_derivation}
\noindent
For the negative log-posterior associated with the probabilistic model of \eqref{eq:EU_IWLP_prb}, we use \eqref{eq:EU_aux_func} and \eqref{eq:IW_nlpr} to define the auxiliary function as
\begin{equation}\label{eq:EU_IW_aux}
    \AuxFnc^{\text{EU-IW}}
    =\\
    \AuxFnc^\text{EU}
    - \log \brc{\prb_\text{IW} \brc{\prm}}.
\end{equation}
The partial derivative of \eqref{eq:EU_IW_aux} with respect to \Zjd{} is given by 
\begin{multline}\label{eq:EU_IWLP_Z_partial}
    \frac{\partial \AuxFnc^{\text{EU-W}}}{\partial \Zjd}
    =
    \frac{\partial \EUAuxFnc}{\partial \Zjd}
    +
    \frac{\partial - \log \brc{\prb_\text{IW} \brc{\prm}}}{\partial \Zjd}
    =
    \frac{\partial \EUAuxFnc}{\partial \Zjd}\\
    +
    \brc{\shdNum-\frDeg}
    \trc{\dirSCM \inv{\estSCM} \dirCov \inv{\estSCM}}
    +
    \brc{\shdNum+\frDeg}
    \trc{\inv{\estSCM} \dirCov},
\end{multline}
which yields the spatial selector update as given by
\begin{multline}\label{eq:EU_IWLP_Z_MU}
    \Zjd
    \xleftarrow[]{}
    \Zjd
    \Bigg[
        \sum_{\freqIdx, \frameIdx}^{\freqNum, \frameNum}
        \frac{\estVjft}{\freqNum \frameNum}
        \trc{\micCovEU \dirCov}
        +
        \frac{\pi \sq{\std_{\text{EU}}} \frDeg}{2}
        \\
        \trc{\dirSCM \inv{\estSCM} \dirCov \inv{\estSCM}}
    \Bigg]
    \inv{\Bigg[
        \sum_{\freqIdx, \frameIdx}^{\freqNum, \frameNum}
        \frac{\estVjft}{\freqNum \frameNum}
        \trc{\mdlCov \dirCov}
        +
        \frac{\pi \sq{\std_{\text{EU}}}}{2}
        \\
        \sqBrc{
            \shdNum \trc{\dirSCM \inv{\estSCM} \dirCov \inv{\estSCM}}
            +
            \brc{\shdNum + \frDeg} \trc{\inv{\estSCM} \dirCov}
        }        
    \Bigg]}.
\end{multline}
Similarly to the algorithm in Sec. \ref{subsec:EU_WLP_derivation}, the localization prior does not directly affect the estimation of spectral parameters, and hence update equations \eqref{eq:EU_Q_MU} - \eqref{eq:EU_H_MU} remain unchanged.
Thus during runtime, updates \eqref{eq:EU_Q_MU} - \eqref{eq:EU_H_MU} and \eqref{eq:EU_IWLP_Z_MU} with re-scaling given by \eqref{eq:Z_MU_rescaling} are sequentially repeated until convergence or some other stopping criterion.

\section{Derivation of update equations with Itakura-Saito divergence}\label{sec:IS_derivations}
\noindent
In this section, we derive the update equations for the Itakura-Saito divergence cost function, in case of ML and MAP estimation with Wishart and Inverse Wishart Localization Priors.
Specifically, the update equations for the existing ML algorithm \cite{munoz2021ambisonics} are presented in Sec. \ref{subsec:IS_ML_derivation}, while derivations of both novel MAP algorithms are presented in Secs. \ref{subsec:IS_WLP_derivation} and \ref{subsec:IS_IWLP_derivation}.

\subsection{Maximum likelihood}\label{subsec:IS_ML_derivation}
\noindent
In case of the negative log-likelihood \eqref{eq:IS_nll}, we define an additional latent component \eqref{eq:lat_cmpII} and the associated auxiliary function \eqref{eq:IS_aux_func}, which are given respectively by
\begin{equation}\label{eq:lat_cmpII}
    \latCmpII
    =
    \mdlCov,
\end{equation}
\begin{multline}\label{eq:IS_aux_func}
    \ISAuxFnc
    =
    \sum
    \inv{\Qjk} \inv{\Wfk} \inv{\Htk} \inv{\Zjd}
    \trc{\micCovIS \latCmpI \inv{\dirCov} \latCmpI}\\
    +
    \log \Bigl|\latCmpII\Bigr|
    +
    \Bigg[
        \vtBrc{\mdlCov}
        -
        \Bigl|\latCmpII\Bigr|
    \Bigg]
\inv{\Bigl|\latCmpII\Bigr|}.
\end{multline}
The partial derivative of the auxiliary function \eqref{eq:IS_aux_func} with respect to \Zjd{} takes the form of 
\begin{multline}\label{eq:IS_Z_partial}
    \frac{\partial \ISAuxFnc}{\partial \Zjd}
    =
    \sum_{\freqIdx, \frameIdx, \cmpIdx}^{\freqNum, \frameNum, \cmpNum}
    \Qjk
    \Wfk
    \Htk
    \trc{\inv{\mdlCov} \dirCov}
    \vtBrc{\mdlCov}
    \inv{\Bigl|\latCmpII\Bigr|}\\
    -
    \inv{\Qjk}
    \inv{\Wfk}
    \inv{\Htk}
    \pow{\Zjd}{-2}
    \trc{\micCovIS \latCmpI \inv{\dirCov} \latCmpI},
\end{multline}
and \eqref{eq:IS_Z_partial} subject to the MU rule \eqref{eq:MU_rule} results in the spatial selector update equation as given by
\begin{equation}\label{eq:IS_Z_MU}
    \Zjd
    \xleftarrow[]{}
    \Zjd
    \frac{
        \sum_{\freqIdx, \frameIdx}^{\freqNum, \frameNum}
        \estVjft \trc{\inv{\mdlCov} \micCovIS \inv{\mdlCov} \dirCov}
    }
    {
        \sum_{\freqIdx, \frameIdx}^{\freqNum, \frameNum}
        \estVjft \trc{\inv{\mdlCov} \dirCov}
    }.
\end{equation}
In case of the remaining model parameters, we provide the following parameter update equations
\begin{equation}\label{eq:IS_Q_MU}
    \Qjk
    \xleftarrow[]{}
    \Qjk
    \frac{
        \sum_{\freqIdx, \frameIdx}^{\freqNum, \frameNum}
        \Wfk \Htk \trc{\inv{\mdlCov} \micCovIS \inv{\mdlCov} \dirCov}
    }
    {
        \sum_{\freqIdx, \frameIdx}^{\freqNum, \frameNum}
        \Wfk \Htk \trc{\inv{\mdlCov} \dirCov}
    },
\end{equation}
\begin{equation}\label{eq:IS_W_MU}
    \Wfk
    \xleftarrow[]{}
    \Wfk
    \frac{
        \sum_{\srcIdx, \frameIdx}^{\srcNum, \frameNum}
        \Qjk \Htk \trc{\inv{\mdlCov} \micCovIS \inv{\mdlCov} \dirCov}
    }
    {
        \sum_{\srcIdx, \frameIdx}^{\srcNum, \frameNum}
        \Qjk \Htk \trc{\inv{\mdlCov} \dirCov}
    },
\end{equation}
\begin{equation}\label{eq:IS_H_MU}
    \Htk
    \xleftarrow[]{}
    \Htk
    \frac{
        \sum_{\srcIdx, \freqIdx}^{\srcNum, \freqNum}
        \Qjk \Wfk \trc{\inv{\mdlCov} \micCovIS \inv{\mdlCov} \dirCov}
    }
    {
        \sum_{\srcIdx, \freqIdx}^{\srcNum, \freqNum}
        \Qjk \Wfk \trc{\inv{\mdlCov} \dirCov}
    },
\end{equation}
readily derived by application of an analogous procedure to that of the spatial selector \cite{munoz2021ambisonics}. Note that during runtime, update equations \eqref{eq:IS_Z_MU} - \eqref{eq:IS_H_MU}, with re-scaling given by \eqref{eq:Z_MU_rescaling}, are sequentially repeated until convergence or some other stopping criterion.
\subsection{Maximum a posteriori with Wishart localization prior}\label{subsec:IS_WLP_derivation}
\noindent
For the the negative-log posterior associated with the MAP probabilistic model of \eqref{eq:IS_WLP_prb}, using \eqref{eq:IS_aux_func} and \eqref{eq:W_nlpr}, we define the following auxiliary function
\begin{equation}\label{eq:IS_WLP_aux_func}
    \AuxFnc^{\text{IS-W}}
    =\\
    \AuxFnc^\text{IS}
    - \log \brc{\prb_\text{W} \brc{\prm}}.
\end{equation}
Since the localization prior does not directly affect the estimation of spectral parameters, the update equations \eqref{eq:IS_Q_MU} - \eqref{eq:IS_H_MU} remain unchanged, while in this section, we consider only the spatial selector update equation. The partial derivative of \eqref{eq:IS_WLP_aux_func} with respect to \Zjd{} is given by 
\begin{multline}\label{eq:IS_WLP_Z_partial}
    \frac{\partial \AuxFnc^{\text{IS-W}}}{\partial \Zjd}
    =
    \frac{\partial \AuxFnc^\text{IS}}{\partial \Zjd}
    +
    \frac{\partial - \log \brc{\prb_\text{W} \brc{\prm}}}{\partial \Zjd}
    =
    \frac{\partial \AuxFnc^\text{IS}}{\partial \Zjd}\\
    +
    \frDeg
    \trc{\inv{\sqBrc{\dirSCM^\text{EU}}} \dirCov}
    +
    \brc{\shdNum - \frDeg}
    \trc{\inv{\estSCM} \dirCov}
    ,
\end{multline}
where
\(
\frac{\partial \AuxFnc^\text{IS}}{\partial \Zjd}
\)
is defined in \eqref{eq:IS_Z_partial}, which yields the spatial selector update equation as given by
\begin{multline}\label{eq:IS_WLP_Z_MU}
    \Zjd
    \xleftarrow[]{}
    \Zjd
    \Bigg[
        \sum_{\freqIdx, \frameIdx}^{\freqNum, \frameNum}
        \frac{\estVjft}{\freqNum \frameNum}
        \trc{\inv{\mdlCov} \micCovIS \inv{\mdlCov} \dirCov}
        \\
        +
        \frDeg
        \trc{\inv{\estSCM} \dirCov}
    \Bigg]
    \inv{\Bigg[
        \sum_{\freqIdx, \frameIdx}^{\freqNum, \frameNum}
        \frac{\estVjft}{\freqNum \frameNum}
        \trc{\inv{\mdlCov} \dirCov}
        \\
        +
        \shdNum \trc{\inv{\estSCM} \dirCov}
        +
        \frDeg \trc{\inv{\dirSCM} \dirCov}
    \Bigg]}.
\end{multline}
\subsection{Maximum a posteriori with Inverse Wishart localization prior}\label{subsec:IS_IWLP_derivation}
\noindent
In case of the negative-log posterior associated with the MAP probabilistic model of \eqref{eq:IS_WLP_prb}, we use \eqref{eq:IS_aux_func} and \eqref{eq:IW_nlpr} to define the auxiliary function
\begin{equation}\label{eq:IS_IWLP_aux_func}
    \AuxFnc^{\text{IS-IW}}
    =\\
    \AuxFnc^\text{IS}
    - \log \brc{\prb_\text{IW} \brc{\prm}}.
\end{equation}
The partial derivative of \eqref{eq:IS_IWLP_aux_func} with respect to \Zjd{} is given by
\begin{multline}\label{eq:IS_IWLP_Z_partial}
    \frac{\partial \AuxFnc^{\text{IS-IW}}}{\partial \Zjd}
    =
    \frac{\partial \AuxFnc^\text{IS}}{\partial \Zjd}
    +
    \frac{\partial - \log \brc{\prb_\text{IW} \brc{\prm}}}{\partial \Zjd}
    =
    \frac{\partial \AuxFnc^\text{IS}}{\partial \Zjd}\\
    +
    \brc{\shdNum-\frDeg}
    \trc{\dirSCM \inv{\estSCM} \dirCov \inv{\estSCM}}
    +
    \brc{\shdNum+\frDeg}
    \trc{\inv{\estSCM} \dirCov},
\end{multline}
which yields the spatial selector update as given by
\begin{multline}\label{eq:IS_IWLP_Z_MU}
    \Zjd
    \xleftarrow[]{}
    \Zjd
    \Bigg[
        \sum_{\freqIdx, \frameIdx}^{\freqNum, \frameNum}
        \frac{\estVjft}{\freqNum \frameNum}
        \trc{\inv{\mdlCov} \micCovIS \inv{\mdlCov} \dirCov}\\
        +
        \frDeg
        \trc{\dirSCM \inv{\estSCM} \dirCov \inv{\estSCM}}
    \Bigg]
    \inv{\Bigg[
        \sum_{\freqIdx, \frameIdx}^{\freqNum, \frameNum}
        \frac{\estVjft}{\freqNum \frameNum}
        \trc{\inv{\mdlCov} \dirCov}\\
        +
        \shdNum \trc{\dirSCM \inv{\estSCM} \dirCov \inv{\estSCM}}
        +
        \brc{\frDeg + \shdNum} \trc{\inv{\estSCM} \dirCov}
    \Bigg]}.
\end{multline}
Similarly to the algorithm in Sec. \ref{subsec:IS_WLP_derivation}, the localization prior does not directly affect the estimation of spectral parameters, and hence update equations \eqref{eq:IS_Q_MU} - \eqref{eq:IS_H_MU} remain unchanged. Thus during runtime, update equations \eqref{eq:IS_Q_MU} - \eqref{eq:IS_H_MU} and \eqref{eq:IS_IWLP_Z_MU} with re-scaling given by \eqref{eq:Z_MU_rescaling} are sequentially repeated until convergence or some other stopping criterion.
\section{Experimental evaluation} \label{sec:experimental_evaluation}
In this section, we conduct a thorough experimental evaluation of the four proposed MAP based algorithms and compare their performance with reference methods.
We consider over-determined, determined, as well as under-determined scenarios, with both instrumental and speech source signals.
In addition, we evaluate the impact of non-ideal, i.e. corrupted DOAs on the performance of the considered algorithms and propose a mixed MAP-ML estimation approach, which achieves more robust performance in scenarios with imprecise localization information.
Finally, binaural audio samples and the implementation of the proposed algorithms are made available\footnote{\url{https://metlosz.github.io/ambisonic_spatially_informed_ntf/}}.

\subsection{Datasets with music and speech signals}
\noindent
The experimental evaluation is primarily based on first-order Ambisonic recordings, since B-format is currently the most accessible and widely used spatial audio format based on the SHD.
Keeping in mind high hardware entry threshold for the Ambisonic technology, i.e. B-format microphone or a spherical array with a number of channels often counted in tens, the expectations concerning finest audio quality seem to be justified.
On the other hand, an opposite tendency is usually observed in the ad-hoc cases, e.g. when the microphones of the immediate devices are spontaneously used to form an array, such that the emphasis is shifted from the perceived sound attributes towards efficiency and robustness.
Therefore, to be able to speculate on the expected performance in realistic conditions, we decided to conduct all experiments with CD audio quality, that is 44.1 kHz sampling frequency, contrary to the reduced sampling rate, e.g. 16 kHz, as is commonly encountered in the literature for speech processing.

To accommodate a wide range of source signals encountered in real-life deployments, we decided to generate four distinct datasets\footnote{The datasets can be obtained upon request.}, three of which contain music signals and one is a collection of speech signals.
We would like to emphasize that in order to obtain difficult, yet realistic testing conditions, the instrumental datsets are generated such that different instruments are playing the same fragment of a musical piece, although the extreme unison cases are not included.
This, however, resulted in non-zero cross-correlation between the source signals, which is due to an overlap of the harmonic frequencies when playing melody in the same key and with a temporal overlap, occurring as a result of grid-like structure of sound source activity periods, e.g. when instruments accentuate the same passage in a piece of music.
The first dataset, hereafter referred to as \textsc{Music I}, contains samples of bass, maracas, percussive instruments with cymbals, piano and string synthesizers, from the song entitled "Scarlett" by courtesy of Eddie Garrido, downloaded from "The Mixing Secrets" website\footnote{\url{https://cambridge-mt.com/ms/mtk/\#FunkyGroovesAndAllThatJazz}}.
The second dataset, hereafter referred to as \textsc{Music II}, contains samples of violin, bassoon, cello, clarinet, double bass, and flute, from Mozart's Symphony No. 40 in g minor, 1st movement, from \cite{vigeant2010multi}.
The third dataset, hereafter referred to as \textsc{Speech}, contains multilingual utterances of nine distinct male and female speakers, assembled from various datasets, among others \cite{ebutech3253}.
The fourth dataset is based on  \textsc{DSD100}\footnote{\url{https://www.sisec17.audiolabs-erlangen.de/\#/dataset}} \cite{SiSEC16}, which contains 100 music tracks of different styles, along with 4 isolated stems per file, namely drums, bass, vocals and others.
Note, that contrary to the other datasets, these 4 stems do not necessarily contain a single instrument each, but rather everything that fits into the specified category.

The experimental audio files are obtained by convolving the source signals with Ambisonic room impulse responses, generated using the image-source method \cite{allen1979image}.
During simulations either 2, 4 or 6 simultaneously active sound sources and a receiver are located inside a 10 x 8 x 4 m room. For the majority of experiments, the reverberation time is of around 250 ms, apart from an experiment in which reverberation time level varies between 250, 500, and 750 ms.
Since the first-order Ambisonic recordings, known as B-format, consist of 4 input signals, the choice of the number of sources enables to investigate the performance in the over-determined case of 2 sources, the determined case of 4 sources, and the under-determined case of 6 simultaneously active sound sources. However, for completeness of evaluation, we also perform an experiment for a varying SH order, ranging from 1, 2 to 3.
In all experiments, the receiver position is chosen randomly, while the minimum distance to walls of 1 m is always kept.
The sound sources are randomly distributed on a sphere concentric w.r.t to the receiver, at a random distance of 1.5 - 2 m and a minimum angular separation of \(45^\circ\) between the sources.
For each combination of the number of sources, 100 audio files of a 5s duration are generated. To ensure the ease of comparison, the geometrical setup is fixed across all datasets, i.e. provided that the same number of sources is considered, the same positions of the receiving microphone array and sources are ensured, while only the type of the source signals differ across the various datasets.
Note, that in experiments involving different number of sources, in case of the \textsc{DSD100} dataset \cite{SiSEC16}, only 2 of the studied combinations are possible, with 2 and 4 sources.
\subsection{Source signal reconstruction and evaluation procedure}
\noindent
Given the estimated parameters \prm{}, we reconstruct the source images \(
    \estSrcVec \in \mathbb{C}^{\shdNum \times 1}
\), i.e. multichannel source signals estimates as seen by the microphone array, using the Multichannel Wiener Filter (MWF)
\begin{equation}\label{eq:mwf}
    \estSrcVec
    =
    \mdlSrcCov
    \inv{\mdlCov}
    \shdCmpSigVec .
\end{equation}

The reconstructed source images are evaluated using commonly used source separation measures, namely the Signal-to-Distortion-Ratio (SDR), Image-to-Spatial-Distortion-Ratio (ISR), Signal-to-Interference-Ratio (SIR), and Signal-to-Artifacts-Ratio (SAR) \cite{raffel2014mir_eval, vincent2007first}.
Throughout the rest of this article, we consider measures averaged over sound sources within each of the experimental file.
Note that since we aim at reconstruction of the source images in a given acoustic environment, the target reference audio files contain source signal together with room reverberation, i.e. as if only the target sound source was active and recorded in a reverberant room.

For the ease of comparison, we ensure that all of the considered algorithms are initialized with the same predefined set of random parameters.
Each algorithm is run for 500 iterations, a number heuristically determined to provide convergence in the considered experimental setups, where by one iteration we mean full set of 4 update equations.
Concerning the choice of the factorization hyperparameters, we follow \cite{nikunen2018multichannel} and set the number of discrete uniformly distributed directions to
\(
    \dirNum = 162
    .
\)
On the other hand, the overall number of NTF spectral components is equal to
\(
    \cmpNum = 25 \cdot \srcNum
    .
\)
Note that in contrast to NMF, spectral components are shared among all \(J\) sources in the proposed NTF based approach, and thus the overall number of components can be significantly lower than for the more popular NMF based algorithms.
We heuristically verified that the sufficient overall number of components to achieve satisfying separation performance amounts to \(\cmpNum = 50, 100, 150\) for scenarios with  \(J = 2, 4, 6\) sources, respectively.

In all cases that include the EU, the we set hyperparameter
\(
    \std_{\text{EU}} = \pow{\pi}{-1/2}
\)
to enable a reduction with \(\pi\) in \eqref{eq:EU_nll}, which should result in an equal contribution of the prior and the inferred knowledge.
Finally, using the ground-truth separated Ambisonic recordings, we optimized the prior hyperparameter \frDeg{}, producing one value per each reverberation time, by following the procedure described in \cite{duong2013spatial}.
The learned values of \frDeg{}, as required by the performed experiments, are presented in Table \ref{tab:frdeg}.
The other prior hyperparameter
\(\epsilon\)
was calculated as a single value characteristic for the considered room, based on the known image-source diffuse and first-order reflection signals, as the mean early-to-diffuse magnitude or power ratio, respectively, in case of the EU and IS.
Furthermore, in all experiments, except where explicitly stated otherwise, the knowledge of the ground-truth DOAs is assumed.

Although the hyperparameters \frDeg{} and
\(\epsilon\)
in this work are assumed to be known, we briefly present the viability of the proposed methods in real-life applications, where the individual signals with subdivision into early and late parts are not usually known.
To this end, we include the results of a preliminary experiment, within which we compare the performance of the proposed EU-WLP in case of the oracle-per-room and the estimated-per-file hyperparameters.
The oracle-per-room hyperparameters \frDeg{} and
\(\epsilon\)
are obtained with the aforementioned procedure, while the per-file estimation is carried out using a simple beamforming approach.
As part of this approach,
\(\epsilon\)
is estimated based on the output of the Plane Wave Decomposition (PWD) beamformer, steered towards the sound sources, as the ratio of the sum of the estimated direct source signals to the residual diffuse signal.
The procedure of estimation of the hyperparameter \frDeg{} is analogous to the aforementioned procedure \cite{duong2013spatial}, with the exception that the ideal source SCM is replaced by the estimate, obtained via the MWF, which is designed based on the PWD estimated source signals.
Note, that a similar procedure is used later in this work as a reference method for the evaluation of sound source separation.
The results of this experiment are given in Table \ref{tab:prelim_nu}, as a mean SDR measure, averaged over all four datasets, for various reverberation time values ranging from 250, 500, to 750 ms.
This informal experiment suggests that even inaccurate per-file estimates are enough to produce results close to the case of oracle-per-room hyperparameters.
However, in the rest of the experimental evaluation, we use the oracle-per-room hyperparameters, in order to eliminate the influence of the estimation method on the results and hence obtain more reliable conclusions about the proposed algorithms.
Other approaches to the estimation of the considered hyperparamters known to literature are based on basic room characteristics \cite{kuttruff2016room, duong2013raport, duong2013spatial, duong2011acoustically}, which can be either known or estimated directly from the microphone signals, such as the reverberation time estimation in \cite{perez2020blind}.
An in-depth pursuit of a robust estimation technique for the hyperparameters \frDeg{} and
\(\epsilon\)
are left as a future work, while for now we focus on the experimental evaluation of the priors, given parameter values.
\begin{table}[!t]
\centering
\caption{The learned values of \frDeg{}, which are used in performed experiments; N/A indicates that the value was not required.}
\begin{tabular}{c|ccc|ccc|}
\cline{2-7}
                         & \multicolumn{3}{c|}{WLP}                                           & \multicolumn{3}{c|}{IWLP}                                          \\ \cline{2-7} 
                         & \multicolumn{1}{c|}{250 ms} & \multicolumn{1}{c|}{500 ms} & 750 ms & \multicolumn{1}{c|}{250 ms} & \multicolumn{1}{c|}{500 ms} & 750 ms \\ \hline
\multicolumn{1}{|c|}{EU} & \multicolumn{1}{c|}{4.7}    & \multicolumn{1}{c|}{6.4}    & 7.3    & \multicolumn{1}{c|}{4.7}    & \multicolumn{1}{c|}{N/A}    & N/A    \\ \hline
\multicolumn{1}{|c|}{IS} & \multicolumn{1}{c|}{4.0}    & \multicolumn{1}{c|}{N/A}    & N/A    & \multicolumn{1}{c|}{4.5}    & \multicolumn{1}{c|}{N/A}    & N/A    \\ \hline
\end{tabular}
\label{tab:frdeg}
\end{table}
\section{Experimental results and discussion}
\subsection{Comparative evaluation of the proposed MAP algorithms}
\noindent
In this section, we evaluate the performance of the proposed MAP based algorithms, i.e. EU-WLP, EU-IWLP, IS-WLP, and IS-IWLP against their Maximum Likelihood counterparts, denoted hereafter as EU and IS, for the Euclidean distance and Itakura-Saito divergence, respectively, in over-determined, determined, and under-determined scenarios using all four datasets.
In addition, we consider a ML approach proposed in \cite{nikunen2018multichannel} which also exploits prior localization knowledge by using a certain Binary Initialization (BI) strategy.
Furthermore, as another reference method, we use a popular state-of-the-art source separation algorithm referred to as the Fast Multichannel Nonnegative Matrix Factorization (FMNMF)\footnote{\url{https://github.com/tky823/ssspy}} \cite{sekiguchi2019fast}. Finally, we compare the aforementioned methods with two classical spatial filters, i.e. Plane Wave Decomposition (PWD) beamformer and its combination with the MWF (explained below).

\begin{table}[!t]
\centering
\caption{Mean SDR, averaged over datasets, as a function of reverberation time, for EU-WLP in case of oracle and estimated values of \frDeg{} and $\epsilon^\text{EU}$.}
\begin{tabular}{c|c|c|c|}
\cline{2-4}
                                               & 250 ms            & 500 ms            & 750 ms            \\ \hline
\multicolumn{1}{|c|}{oracle room-wise [dB]}    & 7.98 \(\pm\) 1.91 & 4.08 \(\pm\) 0.94 & 2.78 \(\pm\) 0.83 \\ \hline
\multicolumn{1}{|c|}{estimated file-wise [dB]} & 8.42 \(\pm\) 2.09 & 4.74 \(\pm\) 1.32 & 3.14 \(\pm\) 0.98 \\ \hline
\end{tabular}
\label{tab:prelim_nu}
\end{table}

Within the BI strategy \cite{nikunen2018multichannel}, the values in close proximity of the \srcIdx{}-th ground-truth DOA are set to a non-zero value for the \srcIdx{}-th source index, while for all of the other directions and source indices the spatial selector is initialized with zeros.
The underlying motivation for the application of the BI is to prevent spatial overlap between the sound sources, although one other advantage can be pointed out, namely the computational speedup, resulting from disregarding computations for the overwhelming number of directions, for which the spatial selector is equal to zero.
The zone within which the values are non-zero is controlled by an angular distance threshold, which in our experiments was set according to \cite{nikunen2018multichannel} as
\(
    22.5^\circ
    .
\)
We extend this approach, and apply it also to the Itakura-Saito divergence based algorithm.

The PWD references are obtained by an application of the beamforming, followed by a re-spatialization stage, where at both steps we use the SH gains for the known DoAs.
The final reconstructed source images are in this case the PWD estimated source signals, as seen by the microphone array.

The aforementioned combination of the PWD and MWF, which we hereafter refer to as MWF, is based on \eqref{eq:mwf}, with the exception that the estimated covariance matrices of the source images \mdlSrcCov{} are based on the output of the PWD beamformer.
The reference source images are in this case reconstructed via the MWF, which is designed based on the source signals estimated using PWD.

The FMNMF algorithm \cite{sekiguchi2019fast} is a fairly recent Non-negative Matrix Factorization based source separation technique which, in contrast to the considered ML Ambisonic NTF, is characterized by the SCMs that are frequency-dependent and are not direction aware, i.e. the knowledge about the microphone array is not exploited.
The fact that the FMNMF algorithm uses direction-unconstrained SCMs enables its straightforward application to any microphone array configuration, including arrays dedicated to Ambisonics.
Since FMNMF is extensively applied in the current literature, it is selected in this work as a good state-of-the-art reference.

\begin{figure*}[!htb]
\centering
\includegraphics[width=0.975\textwidth]{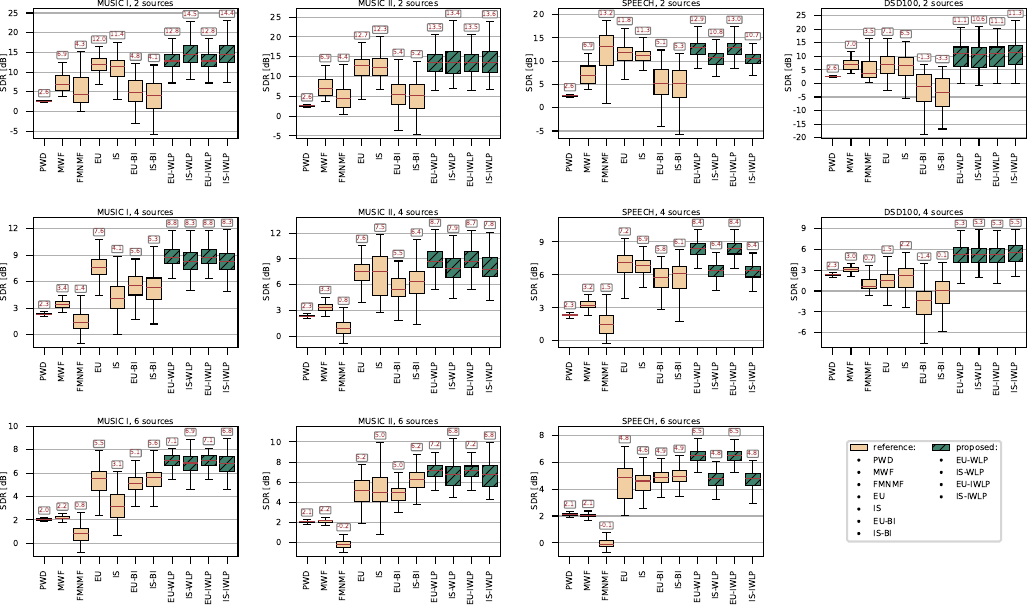}
\caption{Signal-to-Distortion-Ratio (SDR) metrics for all reference and proposed methods. The number of sound sources increases in rows from 2, 4, to 6, while the datasets (\textsc{Music I}, \textsc{Music II}, \textsc{Speech}, \textsc{DSD100}) change in columns. Note that \textsc{DSD100} datasets supports only 2 and 4 sources. Boxes denote 25\% and 75\% quartiles, whiskers mark the minimum and maximum values, while medians are denoted with a red line and their values given above boxplots.}
\label{fig:preliminary-SDR}
\end{figure*}

\begin{figure*}[!htb]
\centering
\includegraphics[width=0.975\textwidth]{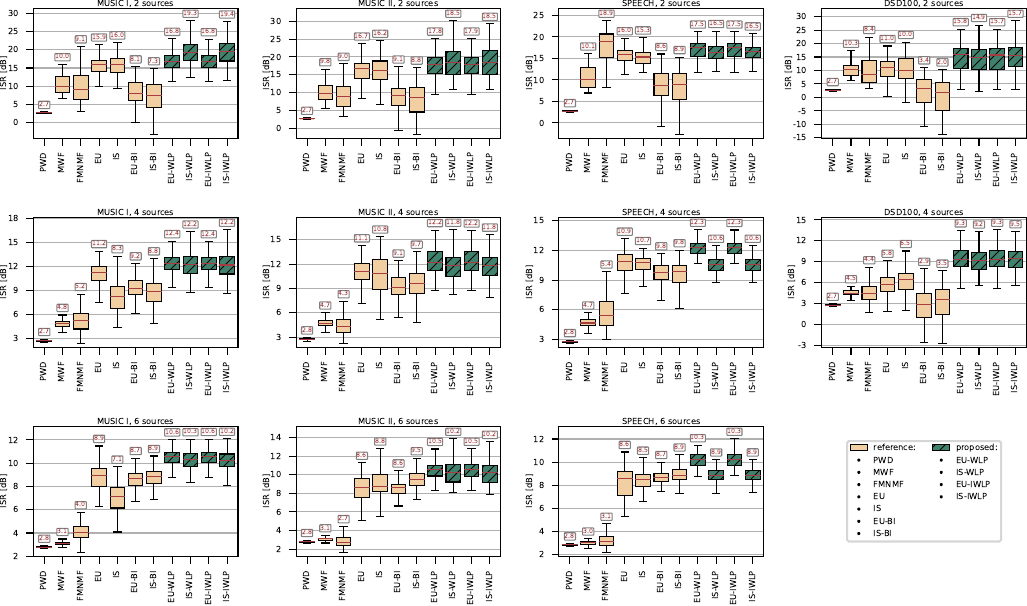}
\caption{Image-to-Spatial-Distortion-Ratio (ISR) metrics for all reference and proposed methods. The number of sound sources increases in rows from 2, 4, to 6, while the datasets (\textsc{Music I}, \textsc{Music II}, \textsc{Speech}, \textsc{DSD100}) change in columns. Note that \textsc{DSD100} datasets supports only 2 and 4 sources. Boxes denote 25\% and 75\% quartiles, whiskers mark the minimum and maximum values, while medians are denoted with a red line and their values given above boxplots.}
\label{fig:preliminary-ISR}
\end{figure*}

\begin{figure*}[!htb]
\centering
\includegraphics[width=0.975\textwidth]{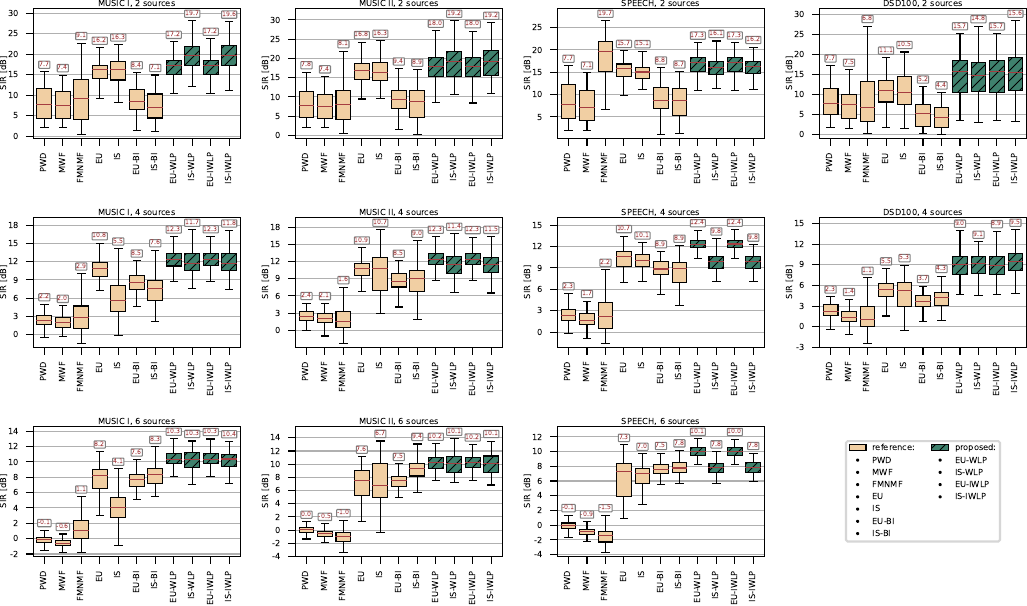}
\caption{Signal-to-Interference-Ratio (SIR) metrics for all reference and proposed methods. The number of sound sources increases in rows from 2, 4, to 6, while the datasets (\textsc{Music I}, \textsc{Music II}, \textsc{Speech}, \textsc{DSD100}) change in columns. Note that \textsc{DSD100} datasets supports only 2 and 4 sources. Boxes denote 25\% and 75\% quartiles, whiskers mark the minimum and maximum values, while medians are denoted with a red line and their values given above boxplots.}
\label{fig:preliminary-SiR}
\end{figure*}

\begin{figure*}[!htb]
\centering
\includegraphics[width=0.975\textwidth]{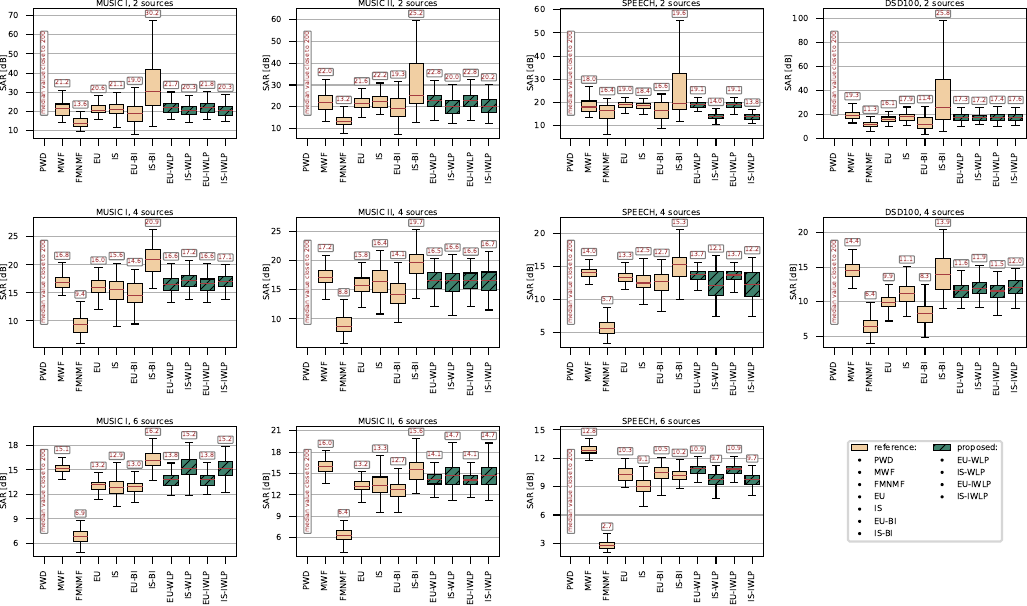}
\caption{Signal-to-Artifacts-Ratio (SAR) metrics for all reference and proposed methods. The number of sound sources increases in rows from 2, 4, to 6, while the datasets (\textsc{Music I}, \textsc{Music II}, \textsc{Speech}, \textsc{DSD100}) change in columns. Note that \textsc{DSD100} datasets supports only 2 and 4 sources. Boxes denote 25\% and 75\% quartiles, whiskers mark the minimum and maximum values, while medians are denoted with a red line and their values given above boxplots.}
\label{fig:preliminary-SAR}
\end{figure*}

\begin{figure*}[!ht]
\centering
\includegraphics[width=0.975\textwidth]{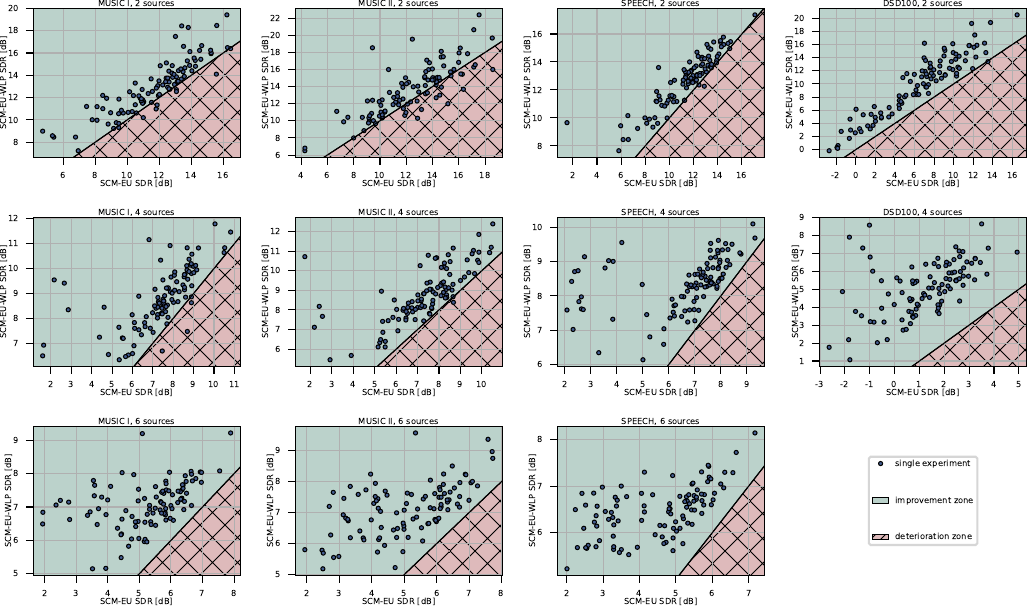}
\caption{File-wise comparison of the input-output mean SDR for the baseline EU and the proposed EU-WLP. The number of sound sources increases in rows from 2, 4, to 6, while the datasets (\textsc{Music I}, \textsc{Music II}, \textsc{Speech}, \textsc{DSD100}) change in columns. Note that \textsc{DSD100} datasets supports only 2 and 4 sources.}
\label{fig:in_out_SDR}
\end{figure*}

Fig. \ref{fig:preliminary-SDR} depicts boxplots of the SDR metric averaged over all sound sources within each experimental audio file, for all 11 compared algorithms, all files from 4 datasets, and for the varying number of sources. 
Interestingly, among the unconstrained ML approaches, namely EU and IS, the EU cost function perform best in each experimental setup.
The EU results are characterized by higher SDR medians, while for the \textsc{Music I} dataset with 4 and 6 sound sources, the 25\% quartile of EU is located over the 75\% quartile of IS.
We hypothesize that the advantage of the EU w.r.t the IS is related to its scale sensitivity, which might be advantageous for spatial features, since we observed on isolated files that the EU is faster (in terms of the number of iterations) to localize the direct sound.

The BI is beneficial in only one experimental setup for the EU algorithm, with the \textsc{Speech} dataset for 6 sources, where the EU-BI SDR distribution has a slightly higher median, with more concentrated values than the baseline EU, therefore indicating better replicability.
Concerning the BI for the IS algorithm, an improvement is observed for the under-determined case of 6 sources for all datasets and for the \textsc{Music I} dataset with 4 sources.
The largest improvement obtained with the IS-BI w.r.t the IS can be observed for the \textsc{Music I} dataset with 6 sources, where the median SDR is significantly higher, while the 25\% quartile of IS-BI is located over the 75\% quartile of the baseline IS.
Although in selected cases some improvement is indeed observed for the BI, in majority of experimental setups this initialization strategy results in a severe degradation of the separation quality, as given by strong deterioration of the median SDR w.r.t. the EU or the IS cost functions.
This is most likely a result of the incompatibility between the BI and the adopted reconstruction and evaluation scheme, i.e. we evaluate the estimated signals with respect to both the direct and the reverberant contributions, whereas the BI primarily extracts the direct sound, as explained in the following.
Since the directions outside a close proximity of the ground-truth DOAs are not taken into account, most of the reflections and late diffuse sound are not accounted for in the SCMs, which becomes apparent when equation \eqref{eq:SCM} is recalled.
This inability of the BI to accurately represent the reverberant sound field is reflected in the SDR values.
Therefore, the BI could be more suited for applications where only the direct sound is desired, e.g. combined with a reconstruction based on plane wave decomposition beamformer followed by a single-channel Wiener postfilter.

The state-of-the-art FMNMF algorithm performs quite well for a low number of uncorrelated sources, such as is the case for \textsc{Speech} with 2 sources. In contrast, it performs poorly for correlated sources, such is the case for \textsc{Music I} and \textsc{Music II} datasets, irrespective on the number of sources. In fact, it consistently achieves the lowest SDR amongst all compared methods for 4 and 6 sound sources encountered in the majority of datasets. Such poor separation performance of the FMNMF algorithm in determined and under-determined scenarios, and in scenarios with correlated sources, can be attributed to having too many degrees of freedom, which results from the lack of additional constrains imposed on the optimization problem. In general, the proposed algorithms perform significantly better than the state-of-the-art FMNMF algorithm, proving the provide robust performance even in difficult determined and under-determined scenarios with correlated sources. In comparison with the reference spatial filters, PWD beamformer and MWF, all proposed algorithms achieve significantly better source separation as indicated by much larger SDR values.

Since in Fig. \ref{fig:preliminary-SDR} the SDR results for the proposed WLP and IWLP do not seem to differ in a significant way, we next consider the results for the MAP estimation more generally, independently of the distribution used to incorporate prior localization knowledge, so that when we refer to EU-MAP it should be read either as EU-WLP or EU-IWLP, and analogously in case of IS.
In case of the EU-MAP, as the number of sources increases, so does the relative improvement obtained w.r.t the EU, irrespectively of the dataset, which is reflected by the growing differences in SDR median.
Similar tendency can be observed for the IS-MAP, although for the \textsc{Music II} dataset this relationship is less pronounced, while for the \textsc{Speech} dataset, MAP based estimation is advantageous only in case of 6 sources.
This would suggest that the improvement obtained with the IS-MAP w.r.t the baseline IS stems from resolving the correlation ambiguity, i.e. the problem of determining if correlation in data is associated with the source signals or if it is connected to the spatial localization.
On the other hand, when only residual correlation is registered between the speech signals, the prior localization knowledge seem to only be advantageous in an extreme under-determined case of 6 speakers.

Overall, in each of the experimental scenarios best SDR results are achieved with one of the proposed MAP solutions.
Since we observe no significant differences in performance between WLP and IWLP, in the remaining part of this article we consider WLP only, as certain parts of the WLP update \eqref{eq:EU_WLP_Z_MU} can be pre-computed, which offers a slight computational advantage over the IWLP.

Figs. \ref{fig:preliminary-ISR}-\ref{fig:preliminary-SAR} present boxplots for the ISR, SIR and SAR metrics averaged over the sound sources within each experimental audio file, for both the EU and the IS cost functions, with and without the WLP.
Based on this extended set of measures, a more insightful analysis of the relations between the ML and the MAP solutions can be attempted, such as verification whether the overall SDR improvement is not a result of an increased attenuation of the undesired sources at the expense of more pronounced artifacts.
In case of the EU cost function, as the number of sources increases, the incorporation of localization information via WLP results in a progressively growing relative improvement for all evaluation measures, which is consistent with the aforementioned tendency observed in Fig. \ref{fig:preliminary-SDR}.
It can be seen that compared to the baseline EU, the proposed EU-WLP allows for a better preservation of the spatial properties, as given by the ISR metric.
Furthermore, compared to the baseline EU, the proposed EU-WLP enables to achieve more attenuation of the undesired sound sources, as given by the substantial improvements in the SIR measure, especially well pronounced for the \textsc{Speech} dataset with 6 sources.
On the other hand, the improved separation and enhanced spatial properties are not compromised by an introduction of notable artifacts, as confirmed by an increase in the SAR score.
This suggests that if the WLP prior is well structured, the proposed EU-WLP is very likely to be beneficial in a wide range of estimation scenarios, without strong risk of a negative influence.

Regarding the IS-WLP, it can be observed that when the spatial localization is not very demanding, such as is the case for 2 sound sources or for the \textsc{Speech} dataset, due to the aforementioned relatively reduced cross-correlation between the source signals, the improvement w.r.t. the IS is not consistent across all measures and experimental scenarios.
In all of the cases that include 2 sound sources a degradation of SAR w.r.t the plain IS can be observed, especially for 2 speakers, where the improvement of ISR and SIR can no longer compensate for the increased amount of artifacts, which is reflected in a noticeably lower SDR median, as shown in Fig. \ref{fig:preliminary-SDR}.
In a 4 speaker scenario, a slight degradation in all measures can also be observed, while with 6 speakers the relative improvement recovers, as the task of spatial localization of 6 sources in an under-determined case with 4 microphones becomes troublesome.
Overall, since the EU-MAP is proven more effective than the IS-MAP in majority of cases, while it also offers a large computational advantage as described in Secs. \ref{subsec:EU} and \ref{subsec:IS}, in the following we shift our focus towards the EU-WLP.

To be able to speculate about the file-wise influence of the proposed EU-WLP w.r.t the baseline EU, i.e. to determine how does the presented statistical SDR improvement relate exactly to each file and experimental setup, we present the input-output mean SDR graphs depicted in Fig. \ref{fig:in_out_SDR}.
In the graph, each dot represents a single file, such that their X-axis coordinates are determined by the mean SDRs obtained with the baseline EU, while the Y-axis coordinates correspond to the mean SDRs obtained with the EU-WLP.
The input-output space is divided by an identity line
\(
    x=y
    ,
\)
which denotes the threshold between the improvement and degradation of the mean SDR, such that the points that lie above this line fall into an improvement zone, while points below this line are in the deterioration zone.

The results depicted in Fig. \ref{fig:in_out_SDR} indicate that with a growing number of sources, the risk of unintentional degradation of the mean SDR is reduced.
In the case of 2 sound sources, some degradation is observed, especially for the \textsc{Music II} dataset, although generally the overwhelming majority of audio files is located in the improvement zone.
For 4 sound sources, only single cases of deterioration can be observed, while the overall distribution slowly starts to flatten out.
Finally, in case of 6 sources, the deterioration zones are empty, while notable and constant relative improvement is observed, irrespectively of the mean input SDR.
Overall, it seems that the proposed EU-WLP can be applied instead of the baseline EU without great risk of relative SDR deterioration.
The application conditions for which incorporation of the proposed localization prior is advantageous include low input SDR, determined and undetermined cases, as well as speech source signals, which suggests that the proposed MAP estimation with the EU-WLP is especially advantageous in challenging acoustic conditions.
\subsection{Performance of the proposed MAP approach for different orders of spherical harmonics and reverberation time levels}
\label{ssec:SH_RT}
\noindent
In this section, we investigate an impact of a varying order of spherical harmonics and reverberation time levels on the performance of the proposed MAP-based source separation. In particular, we perform this study using one of the proposed MAP algorithms, namely the EU-WLP, for a constant number of 4 sources and for all datasets, namely \textsc{Music I}, \textsc{Music II}, \textsc{Speech}, and \textsc{DSD100}.

\begin{figure}[!ht]
    \centering
    \includegraphics[width=0.975\linewidth]{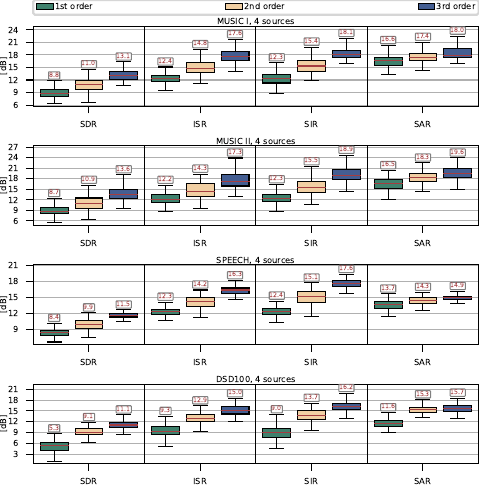}
    \caption{SDR, ISR, SIR and SAR metrics for the proposed EU-WLP algorithm at a varying order of spherical harmonics of 1, 2, and 3, for all four datasets (\textsc{Music I}, \textsc{Music II}, \textsc{Speech}, \textsc{DSD100}) and 4 sources. Boxes denote 25\% and 75\% quartiles, whiskers mark the minimum and maximum values, while medians are denoted with a red line and their values given above boxplots.}
    \label{fig:SH-order}
\end{figure}

\begin{figure}[!ht]
    \centering
    \includegraphics[width=0.975\linewidth]{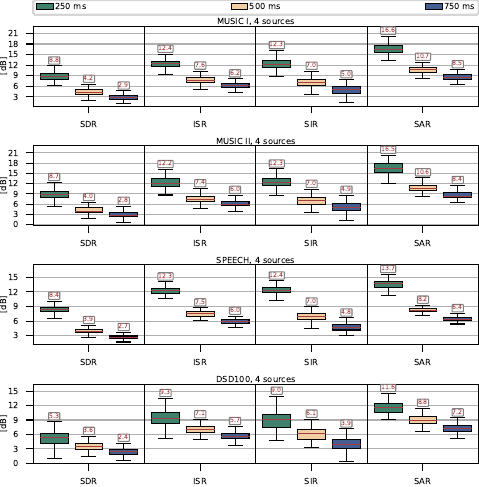}
    \caption{SDR, ISR, SIR and SAR metrics for the proposed EU-WLP algorithm at a varying reverberation time of 250, 500, and 750 ms, for all four datasets (\textsc{Music I}, \textsc{Music II}, \textsc{Speech}, \textsc{DSD100}) and 4 sources. Boxes denote 25\% and 75\% quartiles, whiskers mark the minimum and maximum values, while medians are denoted with a red line and their values given above boxplots.}
    \label{fig:RT60}
\end{figure}

In the first experiment, we consider Ambisonic signals at a varying order of spherical harmonics from the range of 1, 2, and 3. The results of the EU-WLP algorithm in terms of the SDR, ISR, SIR and SAR measures are depicted in Fig. \ref{fig:SH-order}. Across the datasets, all separation performance measures very consistently and significantly increase from the spherical harmonic order of 1, 2, up to 3. For instance, when comparing the first-order and third-order spherical harmonics for the \textsc{DSD100} dataset, SIR increases by a large margin from 9.0 to 16.2 dB, the spatial image of the source is significantly better preserved, while the overall separation measure is more than doubled (SDR increases from 5.3 to 11.1 dB). A consistent improvement in separation performance with an increasing order of spherical harmonics is well to be expected since for the higher number of input channels the considered scenario with 4 sources becomes an over-determined problem. However, an improvement by such a large margin indicates that the proposed MAP approach is capable of achieving remarkable separation of several sources at high-order Ambisonics.

In the next experiment, we investigate separation performance from B-format recordings made at three different reverberation time levels of 250, 500, and 750 ms. The SDR, ISR, SIR and SAR results of the investigated EU-WLP algorithm are shown in Fig. \ref{fig:RT60}. An increase in the reverberation time level clearly causes a notable decrease in all evaluation measures. The largest drop in performance is evident between 250 and 500 ms for the \textsc{Speech} dataset, while a steady and nearly linear decrease in the majority of evaluation measures is observed for the \textsc{DSD100} dataset. Although the presence of late room reverberation makes interference suppression and spatial image preservation more difficult for the proposed algorithm, the overall separation performance achieved by EU-WLP is still satisfactory, even at high reverberation levels.
\subsection{Influence of corrupted DOAs on the performance of the MAP algorithm and of the proposed MAP\textrightarrow{}ML algorithm}
\noindent
This section investigates an impact of increasingly imprecise prior localization information on separation performance of the proposed MAP processing. This study is undertaken for the example EU-WLP algorithm, applied already in Section \ref{ssec:SH_RT}.
To be able to investigate the relationship between the EU-WLP performance and progressively more corrupted DOAs, we introduce the angular distance from the ground-truth direction, aka the angular error, given as
\begin{equation}\label{eq:angular_error}
    \xi^\circ
    =
    \frac{180^\circ}{\pi}
    \arccos
    \brc{
        \tps{\mathbf{d}_\srcIdx} \widehat{\mathbf{d}}_\srcIdx
        \inv{\sqBrc{\vtBrc{\mathbf{d}_\srcIdx} |\widehat{\mathbf{d}}_\srcIdx|}}
    }
    ,
\end{equation}
where
\(
    \mathbf{d}_\srcIdx \in \mathbb{R}^{3 \times 1}
\)
and
\(
    \widehat{\mathbf{d}}_\srcIdx \in \mathbb{R}^{3 \times 1}
\)
denote the ground truth and the corrupted DOA vectors in Cartesian coordinates, respectively.
To obtain the DOA corrupted with a known angular error, equation \eqref{eq:angular_error} needs to be solved for
\(
    \widehat{\mathbf{d}}_\srcIdx
    ,
\)
with a desired value of 
\(
    \xi^\circ
    ,
\)
which results in a set of solutions that form a cone centered around the ground-truth direction.
An intersection of the cone with a unitary sphere determines a constrained set of suitable solutions, the equi-angular error circle, characterized by the desired angular error, yet with undetermined direction.
To produce the final solution, we randomize the exact location of the corrupted DOA on the equi-angular error circle, but with a constraint that the resulting directions 
\(
    \widehat{\mathbf{d}}_\srcIdx
\)
should be at least
\(
    45^\circ
\)
apart for the different sound sources, since a similar threshold is often a part of DOA estimation methods.
Note that with the above formulation, depending on the separation between the ground-truth DOAs and the desired angular error, it is possible that the \srcIdx{}-th corrupted DOA points closer towards a different source than anticipated.
Therefore, to address that problem, we calculate the angular errors w.r.t. the closest ground-truth DOA.
In the remaining parts of this section, we refer to the adopted angular error and DOA corruption strategy as constrained cone error.
The motivation underlying the introduction of this somewhat new metric and procedure is to highlight that our approach is substantially different from the commonly used DOA corruption schemes, based on introduction of an additive noise with gradually increasing power and determining the angular error in a post factum manner.
Since our approach ensures that each experimental file is characterized with exactly the same magnitude of angular error, while the geometrical setup remains identical between the datasets and only varies across different number of sources, we argue that our technique enables to obtain a more reliable and replicable results.
Although the case of uniform errors across sound sources seems very extreme and possibly not very realistic, we note and agree to that compromise, given the aforementioned advantages.

\begin{figure}[!htb]
\centering
\includegraphics[width=\linewidth]{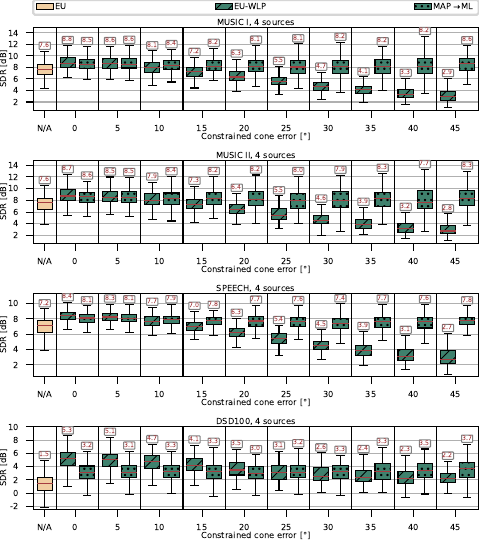}
\caption{SDR results of the proposed EU-WLP algorithm as a function of the constrained cone error for the \textsc{Music I}, \textsc{Music II}, \textsc{Speech}, and \textsc{DSD100} datasets, with 4 sources. Boxes denote 25\% and 75\% quartiles, while medians are denoted with a red line and their values are given above boxplots.}
\label{fig:SDR_angular_error}
\end{figure}

Fig. \ref{fig:SDR_angular_error} presents the distribution of the mean SDR in case of 4 sources, as a function of the constrained cone error for the standard EU, the EU-WLP and a certain combination of EU-WLP and EU, referred to as MAP\textrightarrow{}ML algorithm.
Within the MAP\textrightarrow{}ML, after 450 iterations of the EU-WLP, the estimated parameters \prm{} are used for initialization of the standard ML based EU and 50 more iterations are performed with the EU, as presented in Algorithm \ref{alg}.
The results for the standard EU, with 500 iterations, serve as a bottom-line, such that to uphold the rationale, the proposed solution should perform better than EU.
Once the constrained cone error raises above
\(
    10^\circ
\)
, the performance of the EU-WLP starts to steadily degrade w.r.t. the standard EU.
On the other hand, for the constrained cone error below
\(
    10^\circ
\)
the MAP\textrightarrow{}ML performs slightly worse than EU-WLP, while it seems to be able to uphold its superior performance w.r.t. the standard EU, even when provided with only a rough direction estimate, such as "front left upper corner", which corresponds to
\(
    45^\circ
\)
of constrained cone error.
Although for high values of constrained cone error, the mean SDR results for the MAP\textrightarrow{}ML are more spread out, the median SDR is usually notably greater.
Significant gain in SDR is observed for the constrained cone error larger than \(10^\circ\) for the \textsc{Music I}, \textsc{Music II}, and \textsc{Speech} datasets, while small gain is apparent only at very high constrained cone error values for the \textsc{DSD100} dataset.
Overall, subject to the expected accuracy of the prior localization information, either the EU-WLP or the MAP\textrightarrow{}ML should be applied.

{\LinesNotNumbered
\begin{algorithm}[!ht]
\footnotesize{
\SetKwInput{Parameters}{Parameters}
\SetKwInput{Input}{Input}
\SetKwInOut{Output}{Output}

\Parameters{$\widehat{\mathbf{d}}_\srcIdx$, \srcNum{}, \cmpNum{}, \frDeg{}, $\epsilon$, \dirCov{}}
\Input{$\micCovEU$}
\texttt{initialize \Qjk{}, \Wfk{}, \Htk{}, \Zjd{} with random non-negative values}

\For{\texttt{500 iterations}}{
\texttt{update \Qjk{} with \eqref{eq:EU_Q_MU}}\\
\texttt{update \Wfk{} with \eqref{eq:EU_W_MU}}\\
\texttt{update \Htk{} with \eqref{eq:EU_H_MU}}\\
    \If{\texttt{iteration < 450}}{
        \texttt{update \Zjd{} with \eqref{eq:EU_WLP_Z_MU}}
   }\Else{
        \texttt{update \Zjd{} with \eqref{eq:EU_Z_MU}}
   }
\texttt{rescale \Zjd{} with \eqref{eq:Z_MU_rescaling}}\\
}
\Output{\Qjk, \Wfk, \Htk, \Zjd}
 \caption{MAP\textrightarrow{}ML} 
 \label{alg}
}
\end{algorithm}
}

Next, we investigate the relative performance degradation of the MAP\textrightarrow{}ML w.r.t. the EU-WLP, which is observed in Fig. \ref{fig:SDR_angular_error}, for the lowest values of the constrained cone error. We address the most extreme case of
\(
    0^\circ
    ,
\)
to speculate about the separation performance, should the confidence in accuracy of the prior localization information be underestimated and MAP\textrightarrow{}ML would be improperly chosen instead of EU-WLP.

\begin{figure*}[!htb]
\centering
\includegraphics[width=0.9\textwidth]{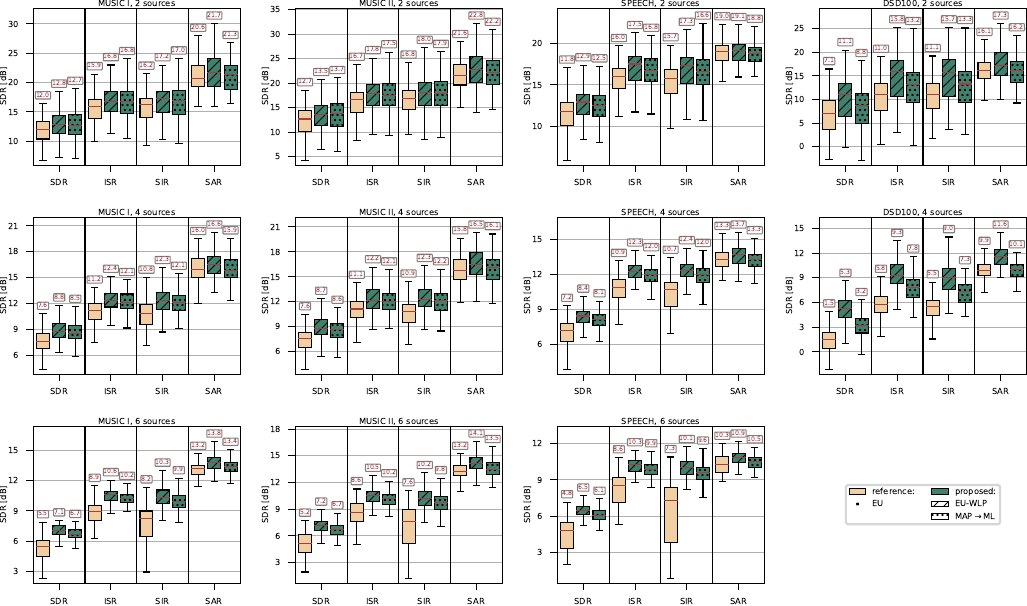}
\caption{SDR, ISR, SIR and SAR results for the standard EU and the proposed EU-WLP and MAP\textrightarrow{}ML algorithms with \(0^\circ\) of constrained cone error. The number of sources increases in rows from 2, 4, to 6, while the datasets (\textsc{Music I}, \textsc{Music II}, \textsc{Speech}, \textsc{DSD100}) change in columns. Note that \textsc{DSD100} datasets supports only 2 and 4 sources. Boxes denote 25\% and 75\% quartiles, whiskers mark the minimum and maximum values, while medians are denoted with a red line and their values are given above boxplots.}
\label{fig:prior_vs_mixed}
\end{figure*}

Fig. \ref{fig:prior_vs_mixed} depicts the SDR, ISR, SIR and SAR measures for the standard EU, the proposed EU-WLP and the proposed MAP\textrightarrow{}ML with oracle localization knowledge.
Consistently across all datasets, the aforementioned SDR performance deterioration of the MAP\textrightarrow{}ML w.r.t. the EU-WLP is not caused by a significant drop in any one single measure, but rather by a slight degradation of all measures.
Although the decrease in evaluation measures somewhat strengthens as the number of sound sources grows, the median SDR in each case is noticeably greater than that of the standard EU.
This suggests that if the confidence in accuracy of the prior localization information is underestimated and MAP\textrightarrow{}ML is accidentally chosen instead of the EU-WLP, there appears to be very little negative impact in terms of the evaluation measures.

\section{Conclusions}\label{sec:conclusions}
\noindent
In this article, we presented a detailed derivation of four algorithms for sound source separation from Ambisonic signals, which enable to efficiently utilize the prior DOA knowledge, in a MAP sense.
The proposed MAP algorithms are based on two cost functions, namely the squared Euclidean distance and the Itakura-Saito divergence, which are combined with two probabilistic prior SCM distributions, namely the Wishart and the Inverse Wishart.
The performance of the proposed MAP algorithms was compared with the existing ML approaches for the spherical harmonic domain, and an interesting combination of the MAP\textrightarrow{}ML methods was introduced for addressing the problem of heavily corrupted localization information.

The derived MAP algorithms were subject to an extensive experimental evaluation, with different number of sources, various types of source signals, and in both the oracle and non-oracle conditions.
The evaluation was based on measures widely used for evaluation of source separation performance, namely, the SDR, ISR, SIR and SAR.
The use cases for the proposed solutions were specified, depending on the acoustic scene and accuracy of prior DOA knowledge.
Overall, in comparison with the existing baseline ML approaches in the SHD, the proposed MAP solution offers a superior separation performance in a wide variety of scenarios, as proven by an analysis of the evaluation measures.

\bibliographystyle{IEEE.bst}
\bibliography{references.bib}
\end{document}